\documentclass[useAMS,usenatbib,usegraphicx]{mn2e}
\usepackage{amsmath,amsfonts,amssymb}
\usepackage[latin1]{inputenc}
\usepackage{times} 
\usepackage{color}     
\usepackage{graphicx}

\usepackage[normalem]{ulem} 
\usepackage{epstopdf}
\usepackage{epsfig}

\newcommand{\ahf}{\textsc{AHF}}
\newcommand{\consistenttree}{\textsc{Consistent Trees}}
\newcommand{\dtree}{\textsc{D-Trees}}
\newcommand{\mergertree}{\textsc{MergerTree}}

\newcommand{\hbt}{\textsc{HBT}}
\newcommand{\jmerger}{\textsc{JMerge}}
\newcommand{\subfind}{\textsc{SubFind}}
\newcommand{\LHaloTree}{\textsc{LHaloTree}}
\newcommand{\sublink}{\textsc{SubLink}}
\newcommand{\treemaker}{\textsc{TreeMaker}}
\newcommand{\velociraptor}{\textsc{VELOCIraptor}}
\newcommand{\ysamtm}{\textsc{ySAMtm}}

\newcommand{\resubmit}[1]{#1}

\newcommand{\Eq}[1]{Equation~\ref{#1}}
\newcommand{\Fig}[1]{Figure~\ref{#1}}
\newcommand{\Figs}[1]{Figures~\ref{#1}}
\newcommand{\Sec}[1]{Section~\ref{#1}}
\newcommand{\Tab}[1]{Table~\ref{#1}}
\newcommand{\Msun}{\mbox{M$_\odot$}}

\newcommand{\halos}{haloes}
\newcommand{\Halos}{Haloes}

\newcommand{\eg}{e.g.}
\newcommand{\SAM}{SA model}
\newcommand{\SAMs}{SA models}

\newlength{\figwidth}
\newlength{\figtable}
\newlength{\figtripple}
\newlength{\resplot}
\title[Merger Trees Comparison]
{Sussing Merger Trees: The Merger Trees Comparison Project}
\author[Srisawat et al.]
{Chaichalit~Srisawat$^1$, Alexander~Knebe$^2$, Frazer~R.~Pearce$^3$, Aurel~Schneider$^1$,
\newauthor
Peter~A.~Thomas$^1$, Peter~Behroozi$^{4}$, Klaus~Dolag$^{7,14}$, Pascal~J.~Elahi$^8$, Jiaxin~Han$^{9,10}$, 
\newauthor
John~Helly$^{10}$, Yipeng~Jing$^{11}$, Intae~Jung$^{15}$, Jaehyun~Lee$^{15}$, Yao-Yuan~Mao$^{4}$, 
\newauthor
Julian~Onions$^3$, Vicente~Rodriguez-Gomez$^{12}$, Dylan~Tweed$^{13}$, Sukyoung~K.~Yi$^{15}$
\\
  $^1$Department of Physics \& Astronomy, University of Sussex, Brighton, BN1 9QH, UK\\
  $^{2}$Departamento de F\'isica Te\'orica, M\'odulo C-15, Facultad de Ciencias, Universidad Aut\'onoma de Madrid, 28049 Cantoblanco, Madrid, Spain\\
  $^3$School of Physics \& Astronomy, University of Nottingham, Nottingham, NG7 2RD, UK\\
  $^{4}$Kavli Institute for Particle Astrophysics and Cosmology \&
  Physics Department, Stanford University, Stanford, CA 94305, USA;\\ 
  \ SLAC National Accelerator Laboratory, Menlo Park, CA 94025, USA\\
  $^{7}$University Observatory Munich, Scheinerstr. 1, 81679 Munich, Germany\\
  $^{8}$Sydney Institute for Astronomy, University of Sydney, Sydney NSW 2016, Australia\\
  $^{9}$Key Laboratory for Research in Galaxies and Cosmology, Shanghai Astronomical Observatory, 80 Nandan Road, Shanghai 200030, China \\
  $^{10}$Institute for Computational Cosmology, Department of Physics, Durham University, South Road, Durham DH1 3LE, UK\\
  $^{11}$Center for Astronomy and Astrophysics, Department of Physics, Shanghai Jiao Tong University, Shanghai 200240, China\\
  $^{12}$Harvard-Smithsonian Center for Astrophysics, 60 Garden Street, Cambridge MA, 02138, USA\\
  $^{13}$Racah Institute of Physics, The Hebrew University, Jerusalem 91904, Israel \\ 
  $^{14}$Max-Planck-Institut f\"ur Astrophysik, Karl-Schwarzschild
Strasse 1, Garching bei M\"unchen, Germany\\
  $^{15}$Department of Astronomy and Yonsei University Observatory, Yonsei University, Seodaemoon-gu Yonsei-ro 50, Seoul 120-749, Republic of Korea
}
\begin{document}
\date{Accepted by MNRAS - \today}

\pagerange{\pageref{firstpage}--\pageref{lastpage}} \pubyear{2013}\volume{0000}

\maketitle

\label{firstpage}

\begin{abstract}
  Merger trees follow the growth and merger of dark-matter \halos\ over
  cosmic history.  As well as giving important insights into the
  growth of cosmic structure in their own right, they provide an
  essential backbone to semi-analytic models of galaxy formation.
  This paper is the first in a series to arise from the {\sc Sussing
    Merger Trees} Workshop in which ten different tree-building
  algorithms were applied to the same set of halo catalogues and
  their results compared. Although many of these codes were similar in
  nature, all algorithms produced distinct results.  Our main
  conclusions are that a useful merger-tree code should possess the
  following features: (i) the use of particle IDs to match \halos\
  between snapshots; (ii) the ability to skip at least one, and
  preferably more, snapshots in order to recover sub\halos\ that are
  temporarily lost during merging; (iii) the ability to cope with (and
  ideally smooth out) large, temporary flucuations in halo mass.
  Finally, to enable different groups to communicate effectively, we defined a
  common terminology that we used when discussing merger trees and we
  encourage others to adopt the same language.  We also specified a
  minimal output format to record the results.
\end{abstract}


\section{Introduction} \label{sec:introduction}

In the era of precision cosmology numerous very large galaxy survey
programmes are either currently underway or in development (just to
name a few, BOSS, PAU, WiggleZ, eBOSS, BigBOSS, DESpec, PanSTARRS,
DES, HSC, Euclid, WFIRST, etc.). The full power of these programmes to
shed light on the nature of dark energy and dark matter can only be
realised if the observational results are compared to theoretical
expectations. Thus the level of precision required can only be
achieved if the theoretical framework is equally well controlled.

Numerical simulations underpin the theoretical predictions for
structure formation and growth. They are required because the
structures that host the galaxies we observe have densities well in
excess of the mean and their growth is highly non-linear. Large
simulations containing billions (soon to be trillions) of tracer
particles have become common in recent years \citep[e.g. Millennium,
DEUS, Bolshoi, MillenniumXXL, Horizon4pi, Jubilee, see][for a recent
review]{Kuhlen_review_2012} and these models cover volumes that are
well matched to aforementioned galaxy surveys covering increasingly
large cosmological volumes. But accurate numerical simulations are not
the end of the story. In order to produce a mock galaxy catalogue the
structures present within these simulated volumes need to be
identified and subsequently populated with galaxies.

By comparing the results obtained for a wide range of halo finding
algorithms, \citet{knebe_haloes_2011} already quantified the errors
introduced during halo identification. This project and its extensions to
the related topics of subhalo detection \citep{onions_2012} and stream
finding \citep{Elahi_2013} are summarised in the review paper
by \citet{knebe_halo_overview_2013}.

Once the set of \halos\ within a cosmological volume have been
reliably identified, the second step is to populate them with
galaxies. This can be done using the information from a single
snapshot by relating the mass of a halo to the number of galaxies it
contains. This is referred to as Halo Occupation Density or HOD
modelling \citep[e.g.~][]{Skibba_2009}. This, however, treats galaxies within
each snapshot independently.  To follow the self-consistent evolution
of galaxies over cosmic time requires information about the growth and
assembly of the \halos\ that host them. The ruleset that determines
how the galaxies contained within these \halos\ form and evolve are
known as semi-analytic models \citep[\SAMs; for a review
  see][]{Bau06}.

\SAMs\ rely on the accuracy of both the individual halo catalogues
themselves as well as the framework that connects the halo catalogues
from different snapshots together. For every object, this framework
forms a tree structure, with many leaves and branches at early times
eventually merging together to form a single trunk that represents the
final galaxy \resubmit{\citep[e.g.~][]{lacey_and_cole_1993,roukema_etal_1997}}. 
The main aim of this paper is to compare and contrast
the tree structures built from a common set of halo catalogues \resubmit{by
ten different tree building algorithms}. We will examine
the accuracy of the trees (how often they link unrelated \halos\
together) and the smoothness of the tree growth. Both can lead to
unrealistic galaxy growth within a \SAM.

The results presented in this paper arise out of the {\sc Sussing
  Merger Tree} workshop, that took place on July 7-12 2013.  In advance
of the workshop, participants were provided with a set of \halos\
(described in Section~\ref{sec:halos} below) and asked to return a
merger tree that linked the \halos\ together over cosmic time in a way
that best represents the growth of cosmic structure.  We allowed
participants to correct errors in their results that arose out of
applying their code to this new data set (e.g.~unusual data format;
periodic boundary conditions) but gave them no feedback in adavnce of
drafting the paper on how their results compared to those of other
participants.

In this paper we use a single set of halo catalogues from a
cosmological box to test the basic properties of the merger trees and
the mass-growth of \halos\ over time.  During the course of the study
presented here it became clear that tree building algorithms are often
intimately tied to the algorithm used to generate the input halo
catalogue, and so in that sense the comparison is not equally fair on
all codes.  While we adhered to this approach in general as it is the
only way to enable an easy comparison between codes, we nevertheless
allowed two codes to modify the halo catalogues
(i.e. \consistenttree\ \& \hbt).  We also allowed algorithms to
convert between inclusive and exclusive particles lists (see
\Sec{sec:terminology}, for a definition) where desired. Future papers
will investigate the effect of changing the halo definition, snapshot
spacing, mass resolution, and eventually the effect on \SAMs.

In what follows, the terminology used throughout the paper will be
specified in Section~\ref{sec:terminology}. Section~\ref{sec:halos}
describes the halo data-set that we use, and Section~\ref{sec:codes}
gives an overview of the various codes that have participated in the
comparison.  We present results on the structure of the resultant
trees in Section~\ref{sec:tree} and of their mass-growth in
Section~\ref{sec:mass}.  Finally, we summarise our results in
Section~\ref{sec:disc}.

\section{Terminology} \label{sec:terminology}

To avoid confusion, it is important that different researchers working
on merger trees speak the same language.  We define here the
terminology used in this paper and would encourage others to adopt the
same definitions:

\begin{itemize}
\item A {\bf halo} is a dark-matter condensation as returned by
  a halo-finder (in our case \ahf). For the purposes
  of other definitions below, we assume that the IDs of the particles
  attributed to each halo by the halo finder are known.
\item \Halos\ may be spatially nested: in that case the outer halo is
  the {\bf main halo} and the other \halos\ are {\bf sub\halos}.  Note
  that the assignment of main halos and sub\halos\ is a function of
  the halo-finder and one can envisage unusual geometries where this
  allocation is not obvious; nevertheless, the picture of
  sub\halos\ orbiting within larger ones ties in with our view of
  cosmic structure and is central to many \SAMs.
\item If particles are allowed to be members of only one halo,
  (i.e.~particles in sub-\halos\ are not included in the particle ID
  list of the main halo, and particles in overlapping \halos\ are
  assigned to just one of the two),
  then the \halos\ are said to be {\bf exclusive}; otherwise they are
  {\bf inclusive} (\ahf\ falls into this latter category).
\item Haloes are defined at distinct {\bf snapshots}. Snapshots correspond to
  particular values of cosmic time and contain the particle IDs, mass,
  location \& velocity for each dark matter particle in the
  simulation.  
\item For two snapshots at different times we refer to the older
    one (i.e. higher redshift) as $A$ and the younger one (i.e. lower
    redshift) as $B$.
\item A {\bf graph} is a set of ordered halo pairs, $(H_A,H_B)$, where
  $H_A$ is older than $H_B$.  It is the purpose of the merger-tree
  codes to produce a graph that best represents the growth of
  structure over cosmic time.  $H_A$ and $H_B$ are usually taken from
  adjacent snapshots, but this is not a requirement as there are
  occasions where \halos\ lose their identity and then reappear at a
  later time.
\item Recursively, $H_A$ itself and progenitors of $H_A$ are {\bf
  progenitors} of $H_B$.  Where it is necessary to distinguish $H_A$
  from earlier progenitors, we will use the term {\bf direct
    progenitor}.
\item Recursively, $H_B$ itself and descendants of $H_B$ are {\bf
  descendants} of $H_A$.  Where it is necessary to distinguish $H_B$
  from later descendants, we will use the term {\bf direct
    descendant}.
\item In this paper we are primarily concerned with {\bf merger trees} for
  which there is precisely one direct descendant for every halo.  Note that
  it is possible for \halos\ near the minimum mass limit to have zero
  descendants: we omit such \halos\ from our analysis.
\item In the case that there are multiple direct progenitors, we
  require that precisely one of these be labelled the {\bf main
    progenitor} -- this will usually be the most massive,
  but other choices are permitted.
\item The {\bf main branch} of a halo is a complete list of main
  progenitors tracing back along its cosmic history.\footnote{We note that,
  for main \halos\ rooted at $z=0$, this main branch might more
  appropriately be called a trunk, but it seems unnecessary to
  introduce a new term for this specific purpose.}
\end{itemize}
Over the course of writing this paper it became clear that there has
been confusion in the past between what we call graphs and merger trees.
Both are interesting in different contexts.  We limit ourselves here
to an investigation of merger trees which are the more relevant as
an input to \SAMs.

\section{Input halo catalogues} \label{sec:halos}

\begin{figure}
\includegraphics[width=\figwidth]{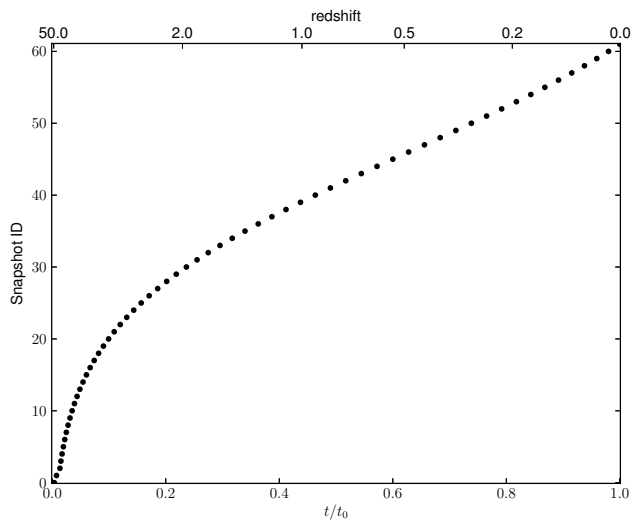}
\caption{Snapshot ID versus time (lower $x$-axis, normalized to
  the present age of the Universe) and redshift (upper $x$-axis).}
\label{fig:snap_info}
\end{figure}

The halo catalogues used for this paper are extracted from 62
snapshots of a cosmological dark matter only simulation undertaken
using the \textsc{Gadget-3} $N$-body code
\citep{springel_cosmological_2005} with initial conditions drawn from
the WMAP-7 cosmology \citep{komatsu_etal_2011}.  We use $270^3$ particles
in a box of comoving width $62.5$ $h^{-1}$Mpc, with a dark-matter
particle mass of $9.31\times10^8h^{-1}$\Msun.  The snapshots are
labelled 0, 1, 2, \ldots, 61 from redshift 50 to redshift 0, as
indicated in \Fig{fig:snap_info}.

The main halo finder used in this paper is \ahf\footnote{The Amiga
  Halo Finder package is publicly available for download from
  \texttt{http://popia.ft.uam.es/AHF}}
\citep{gill_evolution_2004,knollmann_ahf:_2009}.  It locates local
overdensities in an adaptively-smoothed density field as prospective
halo centres. For each of these density peaks the gravitationally
bound particles are determined. Only peaks with at least 20 bound
particles are considered as haloes and retained for further
analysis. The halo mass $M_{200}$ is

\begin{equation} \label{eq:virial_mass_definition}
M_{200} = 200 \rho_{c}(z) \frac{4 \pi}{3} R_{200}^{3},
\end{equation}
where $\rho_{c}(z)$ is the critical density of the Universe as a
function of redshift $z$ and $R_{200}$ is the radius enclosing a mean
density that equals 200 times the critical density.

\ahf\ generates inclusive data sets (i.e.~particles in sub\halos\ are
also included in the main halo).  As an input to the tree-building
codes we provided the list of particle IDs associated with each halo,
alongside information about the (kinetic plus potential) energy,
position and velocity of each particle; we further made available the
full halo catalogue containing, besides the usual mass, position, and
bulk velocity, an abundance of additional information (e.g. energies,
centre offsets, shapes, etc.).

The participants were asked to run their merger tree builders on the
supplied data and return, for each halo, a list of progenitor
\halos\ and (unless the halo was newly-created) the ID of a single
main progenitor.  For the purpose of comparing merger tree algorithms
we restricted participants to use only the information described above
and did not give them access to the raw $N$-body data.  However, they
were allowed to alter the original halo catalogues by adding extra
``fake'' \halos\ and removing some ``unreliable'' \halos\ where they
felt that was appropriate.

\section{Code Descriptions} \label{sec:codes}

In this section we briefly describe, in alphabetical order, the
participating merger tree codes.  Further details of algorithms can be
found in the accompanying references.

The participants were asked to build trees starting from our input
halo catalogues described in Section~\ref{sec:halos}.  One of the
features of a merger tree, as we define it, is that while an object
can have multiple progenitors, only one descendant is allowed. But
many of the algorithms tested did not, in the first instance, produce
a tree. Instead they commonly built graphs that allowed multiple
descendents of a single progenitor halo. To allow consistency and
ensure a fair comparison we required each author to modify their
algorithm to return a tree. Nevertheless, the central process of
linking \halos\ together between snapshots remains and exploring the
various ways of achieving this is the main purpose of this paper.

We note that some of the participating codes required modification in
order to allow them to take as input the \ahf\ halo catalogues that we
used for this comparison project.  To facilitate analysis of the
returned merger trees, we have defined a common, minimal data output
format (described in the Appendix), and this has also required
minor modifications to some of them.

\subsection{Tree Similarity}\label{sec:tree_sim}
\begin{table*}
  \caption{A summary of the features and requirements of merger tree
    algorithms (for details see individual descriptions in the text).
    Columns: (i) Code name; (ii) Particle properties used to produce
    the merger trees; (iii) AHF halo properties used to produce the
    merger trees ($M_{200}$-mass, $\mathbf{r}$-position,
    $\mathbf{v}$-velocity, $V_\mathrm{max}$-maximum rotation speed
    of the halo); (iv)
    the merit function used to estimate descandants; (v)
    the merit function used to estimate the main progenitor; (vi)
    the number of consecutive snapshots used to determine
    descendants/progenitors at each snapshot.
    $\mathcal{M}_1 =N_{A\cap B}^2/(N_{A} N_{B})$, 
    $\mathcal{M}_2 = N_{A\cap B}/N_{B}$,
    $\mathcal{M}_3 = N_{A\cap B}$,
    $\mathcal{M}_4 =\sum_j \mathcal{R}_{(A\cap B)_j}^{-2/3}$,
    $\mathcal{M}_5 = N_{A\cap B}/N_{B}$ for most bound particles only.
}
  \label{tbl:Requirements}
\scriptsize
\begin{tabular}{| l | c | c | c | c | c | c |}
\hline
                   & Particle properties used & AHF halo properties used  & D.Merit Func.\
 & P.Merit Func. & \#Snapshots used \\
\hline
  \consistenttree*  & PID & $M_{200}$, $\mathbf{r}$, $\mathbf{v}$,
  $V_{\rm max}$  & $\mathcal{M}_3$ & Trajectory Est. & 4***    \\
  \dtree           & PID,binding energy      &  --- & $\mathcal{M}_5$ & $\mathcal{M}_5$ & 5***   \\
  \hbt*             & PID,position,velocity &  ---  &  --- & --- & 2  \\
  \jmerger         & --- & $M_{200}$, $\mathbf{r}$, $\mathbf{v}$,
  $V_{\rm max}$   & Trajectory Est.  &  Trajectory Est.  & 2    \\
  \LHaloTree       & PID,binding energy**  &  --- &  $\mathcal{M}_4$ &
  Most massive halo & 3   \\
  \mergertree$^i$      & PID  &  --- &  $\mathcal{M}_1$ & $\mathcal{M}_1$ & 2 \\
  \sublink         & PID,binding energy**  & --- &  $\mathcal{M}_4$\footnotemark  & Most massive history &  3   \\
  \treemaker       & PID  & --- &  $\mathcal{M}_1$ & $\mathcal{M}_1$ & 2  \\
  \velociraptor    & PID  & --- &  $\mathcal{M}_1$ & $\mathcal{M}_1$ & 2  \\
  \ysamtm          & PID  & --- &  $\mathcal{M}_2$ & $\mathcal{M}_2$ & 2  \\
  \hline
  *modify catalogue & **use the distance from halo's   &   &   &
  & ***Users specify but\\
  $^i$ uses the inclusive & centre for this comparison         &
                               & & & these numbers are used \\
particle convention & & & & & for this comparison
\end{tabular}
\normalsize
\end{table*}

\footnotetext{The latest version of \sublink\ uses the index of ranking of $-1$ rather than $-2/3$ which is used in this comparison.}

\begin{figure}
  \centering
  \vspace*{-0.5cm}\hspace*{-0.9cm}
  \includegraphics[width=1.2\linewidth,height=0.8\linewidth,trim={0 4cm 0 0},clip]{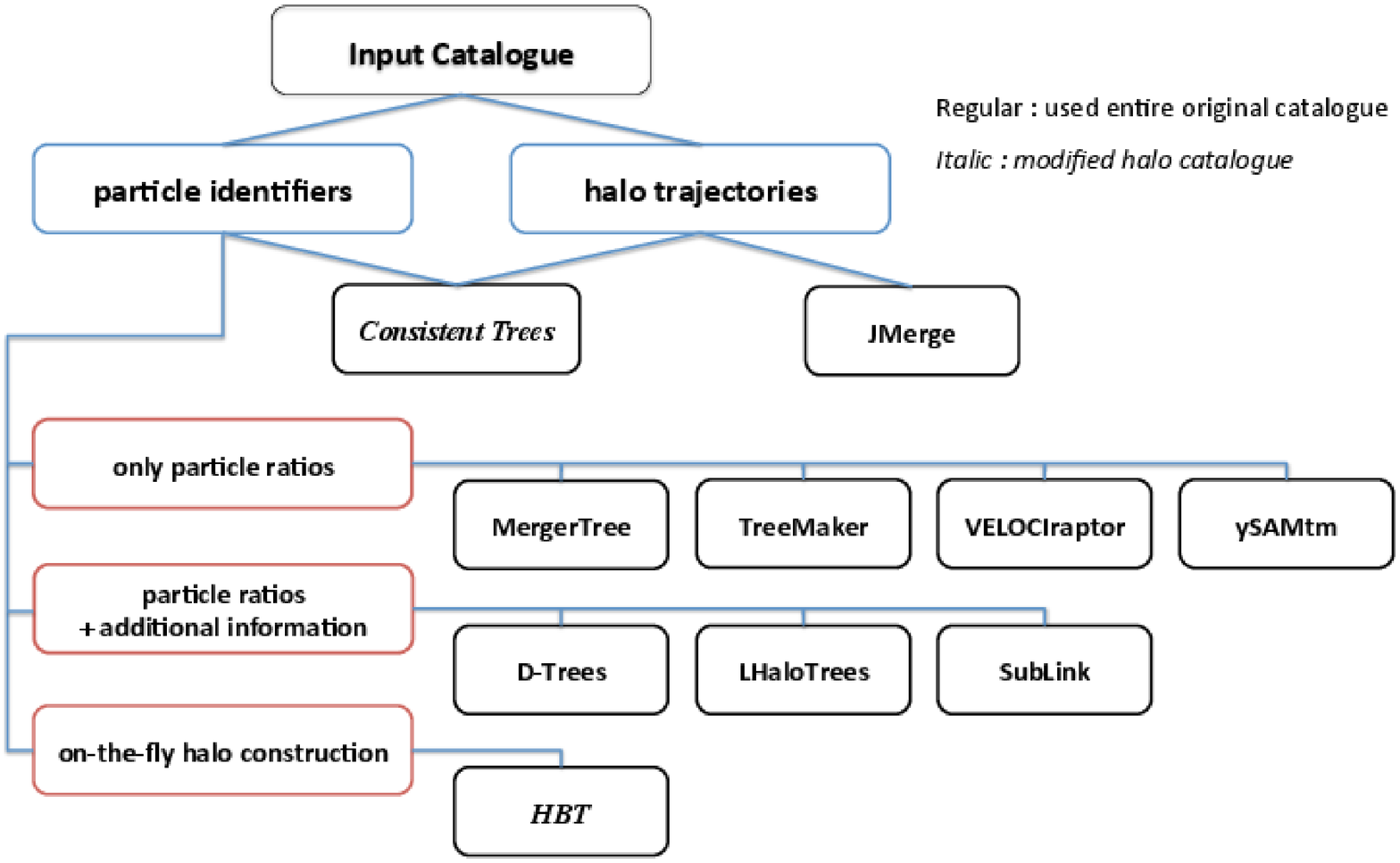}
  \caption{A summary of the main features and requirements of the
    different merger tree algorithms.  For details see the individual
    descriptions in the text.}
\label{fig:overview}
\end{figure}

As a lot of methodology is similar across the various codes used here,
we try to capture the main features and requirements in
\Fig{fig:overview} and Table~\ref{tbl:Requirements}. Note that only a
single code doesn't use particle IDs to link \halos\ between
snapshots: that potentially makes it more widely applicable to legacy
data but leads to problems with misidentification of \halos, as will
be seen later in \Sec{sec:tree} below.

Many tree-codes make use of a merit function
\begin{equation}
    \mathcal{M}(H_A,H_B) = f(N_A,N_B,N_{A\cap B}),
\end{equation}
where $N_A$ and $N_B$ are the number of particles in \halos\ $H_A$ and
$H_B$, respectively, and $N_{A\cap B}$ is the number of particles that
are in both $H_A$ and $H_B$, or
\begin{equation}
    \mathcal{M}(H_A,H_B) = f(\mathcal{R}_{A\cap B}),
\end{equation}
where $\mathcal{R}_{A\cap B}$ is the ranking
(decreasing binding mass or increasng halocentric radius) of particles
that are in both $H_A$ and $H_B$.  Such a function aims at identifying
the most likely progenitor/descendant of a given halo. A few of them
use additional information such as, for instance, the binding energy
of the particles, properties of the \halos\ or information about the
snapshot times.

\subsection{\consistenttree\ (Mao \& Behroozi)}

The \consistenttree\ algorithm \citep{Behroozi_etal_2013} first
matches \halos\ between snapshots by identifying descendant \halos\ as
those that have the maximum number of particles from a given
progenitor halo. It then attempts to clean up this initial guess by
simulating the gravitational bulk motion of the set of \halos\ given
their known positions, velocities, and mass profiles as returned by
the halo finder.  From \halos\ in any given simulation snapshot, the
expected positions and velocities of \halos\ at an earlier snapshot
may be calculated.  In some cases, obvious inconsistencies arise
between the predicted and actual halo properties, such as missed
satellite \halos\ (\eg\ satellite \halos\ which pass too close to the
centre of a larger halo to be detected) and spurious mass changes
(\eg\ satellite \halos\ which suddenly increase in mass due to
temporary miss-assignment of particles from the central halo).  These
defects are repaired by substituting predicted halo properties instead
of the properties returned by the halo finder.  If a halo has no
descendant a merger is assumed to have occurred with the halo exerting
the strongest tidal field across it, unless no such suitable halo
exists in which case the halo is presumed to have been spurious and
this branch is pruned from the merger tree. This process helps to
ensure accurate mass accretion histories and merger rates for
satellite and central \halos; full details of the algorithm as well as
tests of the approach may be found in \cite{Behroozi_etal_2013}.

\subsection{\dtree\ (Helly)}

The D-Trees algorithm (Jiang et al., in preparation) is designed to
work with the \subfind\ group finder, which (like \ahf) can
occasionally fail to detect \halos\ or sub\halos\ for one or more
snapshots. It therefore allows for the possibility that descendants
may be identified more than one snapshot later. Descendants are
identified by following the most bound ``core'' of each group --
i.e.~those particles with the lowest total energy.

To find the descendant at snapshot $B$, of a group which exists at an
earlier snapshot, $A$, the following method is used. For each group
containing $N_p$ particles the $N_\mathrm{link}$ most bound particles are
identified, where $N_\mathrm{link}$ is given by
\begin{equation}
N_\mathrm{link} = \min(N_\mathrm{linkmax}, \max(f_\mathrm{trace} N_{p},
N_\mathrm{linkmin}))
\end{equation}
with $N_\mathrm{linkmin} = 10$, $N_\mathrm{linkmax} = 100$ and
$f_\mathrm{trace}=0.1$. 
Descendant candidates are those groups at snapshot $B$ that received
at least one of the $N_\mathrm{link}$ most bound particles from the earlier
group. If any of the descendant candidates received a larger
fraction of their $N_\mathrm{link}$ most bound particles from the progenitor
group than from any other group, then the descendant is chosen from
these candidates only and the group at snapshot $A$ will be designated
the main progenitor of the chosen descendant; otherwise all candidates
are considered. The descendant of the group at snapshot $A$ is taken
to be the remaining candidate which received the largest fraction of
the $N_\mathrm{link}$ most bound particles of the progenitor group. For each
group at snapshot $B$, this method identifies zero or more progenitors
of which at most one may be a main progenitor. Note that it is not
guaranteed that a main progenitor will be found for every group.

If a group is not found to be the main progenitor of its descendant,
this may indicate that the group has merged with another group and no
longer exists in the simulation. However, it is also possible that the
group finder has simply failed to identify the object at the later
snapshot. In order to distinguish between these cases it is necessary
to search multiple snapshots.

For each snapshot $A$ in the simulation descendants are identified at
later snapshots in the range $A+1$ to $A+N_\mathrm{step}$ using the method
described above. For each group at snapshot $A$ this gives up to
$N_\mathrm{step}$ possible descendants. One of these descendants is picked
for use in the merger trees as follows: if the group at snapshot $A$
is the main progenitor of one or more of the descendants, the earliest
of these descendants that does not have a main progenitor at a
snapshot later than $A$ is chosen. If no such descendant exists, the
earliest descendant found is chosen irrespective of main progenitor
status.

This results in the identification of a single descendant for each
group, which may be up to $N_\mathrm{step}$ snapshots later. Each
group may also have up to one main progenitor which may be up to
$N_\mathrm{step}$ snapshots earlier.

\subsection{\hbt\ (Han, Jing)}

The Hierarchical Bound Tracing (\hbt) algorithm \citep{han_etal_2012}
is a tracking halo finder in the sense that it uses information from
earlier snapshots to help derive the latest halo catalogue. As such it
naturally builds a merger tree. Starting from high redshift, main \halos\
are identified as they form. The particles contained within
these \halos\ are then followed explicitly through subsequent snapshots,
generating a merger tree down to main halo level at the first stage. 
To extend the merger tree down to subhalo level,
\hbt\ continues the tracing of merged branches, identifying the set
of self-bound particles that remain for every progenitor halo. These
self-bound remnants are defined as descendant haloes of their
progenitors. With this kind of tracking, each halo has at most
one progenitor, which defines its main branch. The main branch
extends until the number of particles contained in the bound halo
remnant drops below 20 particles.  When this occurs a final tracking
step is undertaken to determine which halo it has fallen into, adding
minor branches to the tree.

The major challenge in this method is to robustly track
haloes over long periods, and \hbt\ has been specifically tuned
to achieve this. In addition, the merging hierarchy among progenitor
haloes is utilized to efficiently allow satellite-satellite mergers
or satellite accretion inside satellite systems.

Note that \hbt\ is not designed to be a general purpose treebuilder
for external halo catalogues. To generate the trees used in this
paper, \hbt\ was run using only the main haloes from the supplied
catalogue as described in \Sec{sec:halos} as input. It then outputs
its own list of \halos\ and calculates the relevant properties for
them, as well as returning the merger tree built on top of these
\halos.

\hbt\ outputs exclusive halos. In order to give a mass which matches
that of \ahf\ halos as closely as possible, for each halo, we first
calculate an 'exclusive' mass according to
\Eq{eq:virial_mass_definition} using only particles from the halo
itself. Then we add to each halo the exclusive mass of all its
subhaloes, to give an 'inclusive' mass, which we use throughout this
paper.

\subsection{\jmerger\ (Onions)}

The \jmerger\ algorithm constructs a merger tree purely from aggregate
properties (the position, centre-of-mass velocity and mass) of the
\halos\ identified by a halo finder (i.e. it does not require the
individual particle positions or particle IDs). It compares halo
catalogues from two snapshots separated by a known time interval. For
the two sets of \halos\ at times $A$ and $B$, a new position is
calculated for the centre of each halo by moving the $A$ \halos\
forward in time by half the timestep, and the $B$ \halos\ backwards by
half the timestep \resubmit{assuming that they are moving at constant velocities}.
Then, starting from the most massive halo and
working towards smaller masses, for each halo in $A$, a best match on
position is found to a halo in $B$, together with constraints on the
allowed change in mass and maximum circular
velocity. Mass is allowed to shrink by a factor of up to 0.7, and to
grow by a factor of up to 4.  The search distance is limited to twice
the radius at which the enclosed density is 200 times the background
density plus four times the distance the halo has moved during the
timestep.  At this stage, each halo in $B$ can only be claimed once.
This process attempts to trace \halos\ growing over time.

For those \halos\ that do not find an unclaimed descendant in $B$, two
other processes are implemented.  Firstly, mergers are accounted for
by finding so far unmatched \halos\ at time $A$ that can accrete onto
$B$ targets already accounted for, whilst still limiting the total
mass of the direct progenitors of each descendant to less than 1/0.7
times its mass.  Secondly, \halos\ that cannot find a suitable match are
deemed to be numerical artifacts and are pruned from the tree.

\subsection{\LHaloTree\ (Dolag)}

L-HaloTree was the first merger-tree algorithm to construct trees
based on sub\halos\ instead of main halos.  The LHaloTree algorithm is
described in the supplementary information of
\citet{springel_millennium_2005} and the reader is referred there for
further details. In short, to determine the appropriate descendant,
the unique IDs that label each particle are tracked between
outputs. For a given halo, the algorithm finds all halos in the
subsequent output that contain some of its particles. These are then
counted in a weighted fashion, giving higher weight to particles that
are more tightly bound in the halo under consideration, as listed in
\Tab{tbl:Requirements}, and the one with the highest count is selected
as the descendant.  In this way, preference is given to tracking the
fate of the inner parts of a structure, which may survive for a long
time upon infall into a bigger halo, even though much of the mass in
the outer parts can be quickly stripped.

To allow for the possibility that halos may temporarily disappear for
one snapshot, the process is repeated for Snapshot~$n$ to
Snapshot~$n+2$.  If either there is a descendant found in
Snapshot~$n+2$ but none found in Snapshot~$n+1$, or, if the descendant
in Snapshot~$n+1$ has several direct progenitors and the descendant in
Snapshot~$n+2$ has only one, then a link is made that skips the
intervening snapshot.

\subsection{\mergertree\ (Knebe)}

The \mergertree\ routine forms part of the publicly available Amiga
halo finder (\ahf) package. It is a simple particle correlator: it
takes two particle ID lists (ideally coming from an \ahf\ analysis)
and identifies for each object in list\,$B$ those objects in list\,$A$
(at the previous snapshot) with which there $N$ or more particles in
common ($N=10$ for this comparison). Despite its name, therefore, it
produces a graph mapping the connections between objects rather than a
tree, as each halo can have multiple descendants.

\mergertree\ also identifies a unique main progenitor
for each object in list\,$B$ as found in list\,$A$. It achieves this by 
maximising a merit function (as shown in \Tab{tbl:Requirements})
This has proven
extremely successful \citep{Klimentowski_etal_2010, Libeskind_2011,
  Knebe_galaxies_2013}. The code can hence not only be used to trace a
particular object backwards in time (or forward, depending on the
temporal ordering of files\,$A$ and $B$), but also to cross-correlate
different simulations (e.g. different cosmological models run with the
same phases for the initial conditions).

To create an actual tree, we need to ensure that each halo has a
unique descendant. This is guaranteed by running \mergertree\ in a
novel mode that applies the same merit function in both directions
when correlating two files. In practice this links \halos\ that share
the largest fraction of particles between the two snapshots as well as
forcing a choice between multiple possible descendants (of which now
only the one maximising the merit function in the direction
$A\mapsto B$ is kept). The use of a merit function also eliminates
any need for all the particles in the input halo catalogues to only
belong to a single object: $\mathcal{M}_{A_i B_j}$ automatically takes care of particles
that have been assigned to multiple objects.

\subsection{\sublink\ (Rodriguez-Gomez)}

\sublink\ (Rodriguez-Gomez et al., in prep.) constructs merger trees
at the subhalo level. A unique descendant is assigned to each subhalo
in three steps. First, descendant candidates are identified for each
subhalo as those subhalos in the following snapshot that have common
particles with the subhalo in question. Second, each of the descendant
candidates is given a merit function specified in \Tab{tbl:Requirements}.
Third, the unique descendant of the subhalo in question is the 
descendant candidate with the highest merit function.

Sometimes the halo finder does not detect a small subhalo that is
passing through a larger structure, because the density contrast is
not high enough. \sublink\ deals with this issue in the following
way. For each subhalo from snapshot $S_n$, a 'skipped descendant' is
identified at $S_{n+2}$, which is then compared to the 'descendant of
the descendant" at the same snapshot. If the two possible descendants
at $S_{n+2}$ are not the same object, we keep the one obtained by
skipping a snapshot since, by definition, it has the largest score at
$S_{n+2}$.

Once all descendant connections have been made, the main progenitor of
each subhalo is defined as the one with the 'most massive history'
behind it, following De Lucia \& Blaizot (2007). This information is
rearranged into fully-independent merger trees.

\subsection{\treemaker\ (Tweed)}

The \treemaker\ algorithm was developed for the \SAM\
{\tt GalICS} (Galaxies in Cosmological Simulations)
\citep{Hatton2003}. It was first used on Friends-of-Friends
\halos\ \citep{Davis1985}, and later applied to main \halos\ and
sub\halos\ extracted from a cosmological simulation with the {\tt
AdaptaHOP} group finder \citep{Aubert2004,Tweed2009}.  The code
associates \halos\ from two consecutive time steps, listing all
progenitors (including particles accreted from the background) and
descendants (multiple descendants being allowed even if particles lost
to the background are ignored). Here ``background'' refers to particles
not in any halo at the current time. This first step is completed by
using the particle IDs as tracers to identify \halos. Under our scheme
a particle can only belong to one single halo at a given step, meaning
a particle in a subhalo belongs only to that subhalo and not to any
enclosing halo.

In order to create a ``usable'' merger tree a simplification stage is
required. Exactly one descendant per halo is
selected and the list of progenitors updated to reflect this
selection. Selecting this unique descendant requires the use of a
merit function. The first versions of \treemaker\ used a shared
merit function. For this study, we tested various modifications of this
selection, but all gave similar results.  We therefore include in this
paper only the normalised merit function $\mathcal{M}_1$ as shown in
\Tab{tbl:Requirements}.

\subsection{\velociraptor\ (Elahi)}

The halo merger tree algorithm used in \velociraptor\ is based on a
particle correlator: that is the algorithm compares two (or more) {\em
  exclusive} particle ID lists and produces a catalogue of matches for
each object in each list. Specifically, for each object $i$ in
catalogue $A$, the algorithm finds all objects $j$ in catalogue $B$
that share particles, and calculates the strength of each connection
using the merit function $\mathcal{M}_1$ as shown in
\Tab{tbl:Requirements}.
The search for connections is done in both
directions. Any connection with a merit function within Poisson
fluctuations, $\mathcal{M}_{A_{i}B_{j}}\leq 1/(N_{A_{i}}N_{B_{j}})$,
is ignored. The connection that maximises $\mathcal{M}$ for
$A\rightarrow B$ is deemed the unique descendant (note that the
orginal code returned a graph that did not enforce this requirement of
uniqueness). This approach is used as particle ID lists produced by
\velociraptor\ contain not only particles belonging to bound
(sub)\halos\ but also those in physically diffuse tidal
debris. Consequently, tracking object centres or weighting particles
by a measure of how bound they are is meaningless. Note that tidal
debris candidates, due to their physically diffuse nature, can be
artificially fragmented into several \velociraptor\ groups. For
example, a single bound (sub)halo identified at time $A$ is found to
be the progenitor of several tidal debris fragments at time
$B$. Matching $B\rightarrow A$, the fragments identify the (sub)halo
as the primary progenitor, however, the (sub)halo will identify the
largest tidal fragment as its primary descendant. For proposes the of
a this paper, the other fragments are ignored. However, in the general
merger graph produced by \velociraptor, these fragments are
flagged as secondary descendants if fragment shares $\geq5\%$ of
particles with the primary progenitor.

\subsection{\ysamtm\ (Jung, Lee, Yi)} \label{sec:ySAMtm}

The tree-making algorithm \ysamtm\ (Jung et al., in preparation) was
developed to build dark matter halo merger trees for the semi-analytic
model {\tt ySAM} \citep{lee_2013}. It uses the particle information
from two snapshot files or the particle IDs and locations from a
pre-calculated halo catalogue. First the `shared mass', the mass
contribution of all progenitor \halos\ to each descendant halo, is
calculated.  At this stage, particles are matched between \halos\ in
the two snapshots by using the particle IDs. Individual particles are
only included in a single halo or subhalo and are not listed as
members of the host halo of the subhalo. Secondly, in order to convert
our graph into an actual tree that could be used by semi-analytic
models, we define a unique descendant halo of each progenitor halo by
determining which descendant halo has the most shared mass among all
descendants of the progenitor halo, unless there exists a smaller halo
which receives a larger fraction of its mass from the same
progenitor. In this case we determine that the smaller one is the most
likely descendant halo of the progenitor even if its shared mass is
not the largest amongst all the descendants. This avoids defining the
smaller descendant halo as a newly-formed halo when it contains many
particles that were members of an existing halo in the previous
snapshot. This process creates a true tree where one descendant halo
can have multiple progenitor \halos, while each progenitor halo has a
unique descendant halo. Among those progenitors, the main progenitor
is determined by maximising the merit function $\mathcal{M}_2$ in
\Tab{tbl:Requirements}.

\section{Tree structure}\label{sec:tree}

In this section we look at the structure/geometry of trees. This
includes a comparison of measurable quantities like the tree-length
along the main branch, the tree-branching at every step, and the
general consistency of the tree (i.e.~possible mis-identification of
descendants).

\subsection{Length of main branches} \label{sec:length}

\begin{figure}
  \centering
  \includegraphics[width=1\linewidth]{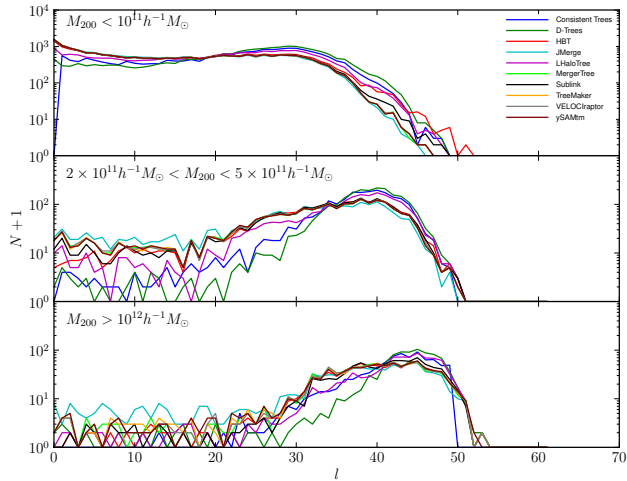}
  \caption{The length of the main branch for \halos\ identified at
    $z=0$ (Snapshot 61). The ordinate is $l=61-S$, where $S$ is the
    snapshot number at the high-redshift end of the main branch.  The
    upper, middle and lower panels show the halo mass ranges at $z=0$,
    as indicated in the panel, which correspond to roughly $<100$,
    200-500 and $>1000$ particles respectively.}
\label{fig:link_depth_mass}
\end{figure}

The most basic requirement of a tree-building code is to trace
\halos\ back in time. The length of the main branch gives a measure of
how long single \halos\ can be followed through the complicated merger
history of structure formation.  \Fig{fig:link_depth_mass} shows the
number, $N$, of $z=0$ \halos\ that have main branches extending for a
given number of snapshots, $l$, for all \halos\ within three different
mass-ranges: \halos\ with $M_{\rm{200}} < 10^{11}$\,$h^{-1}\Msun$
(less than $\sim 100$ particles) are shown in the top panel,
$2\times10^{11}\,h^{-1}\Msun < M_{\rm{200}} <
5\times10^{11}\,h^{-1}\Msun$ in the middle panel, and $M_{\rm{200}} >
10^{12}\,h^{-1}\Msun$ (more than $\sim 1000$ particles) in the bottom
panel.

Large \halos\ (bottom panel of \Fig{fig:link_depth_mass}) tend to have
long main branches with $l=30$--$50$, which is in agreement with the
picture of bottom-up structure formation, where larger objects form
through repeated mergers of smaller ones.

As one moves to smaller \halos\ the proportion of short branches
increases. For $M_\mathrm{200}<10^{11}\,h^{-1}$\Msun\ the number of
main branches per length is roughly constant from $l=0$ until about
$l=30$ (corresponding to $z\approx5$) and only drops to zero beyond
$l\approx50$ ($z\approx10$).  Thus, even in a hierarchical structure
formation scenario, dwarf-sized \halos\ {\it that survive to the
  current day} have a wide variety of formation times.

One oddity in \Fig{fig:link_depth_mass} is that most of
the tree codes find a few large \halos\ with very short main branches
which is in contradiction to the common picture of structure
formation. Further investigation of these branches show that they are
either truncated due to a non-identification by the halo finder, or
are due to an error in the halo assignment of the tree building codes.

\begin{figure*}
  \centering
  \includegraphics[width=1\figtable]{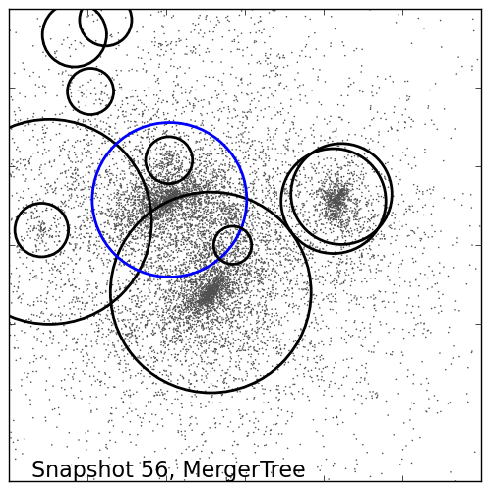} 
  \includegraphics[width=1\figtable]{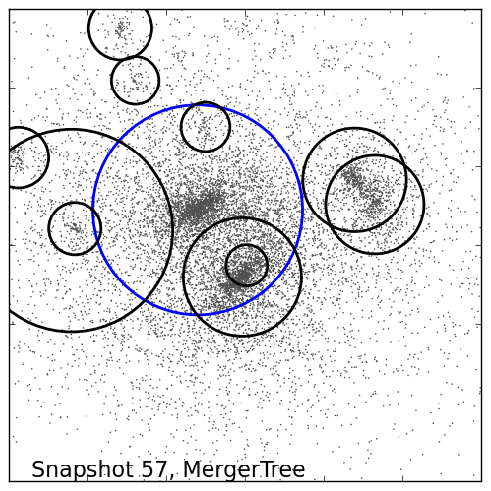}
  \includegraphics[width=1\figtable]{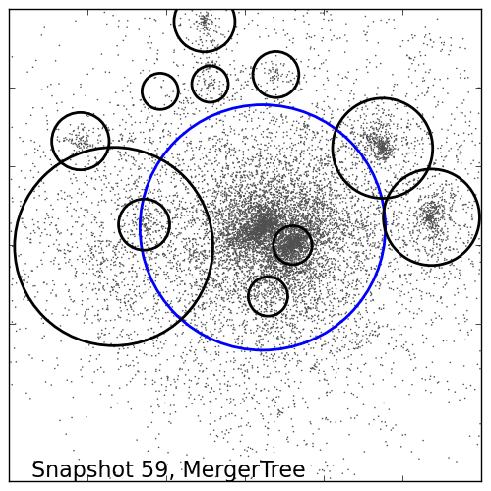} 
  \includegraphics[width=1\figtable]{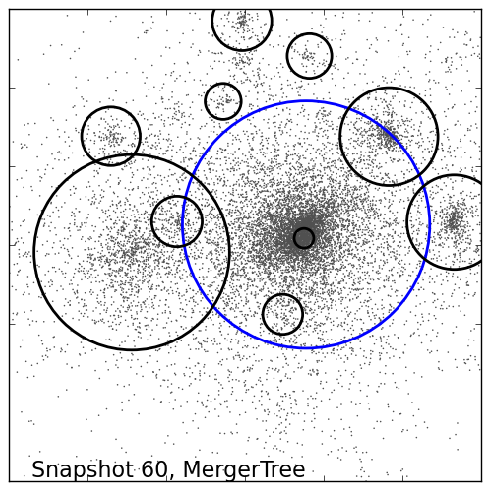} 
  \includegraphics[width=1\figtable]{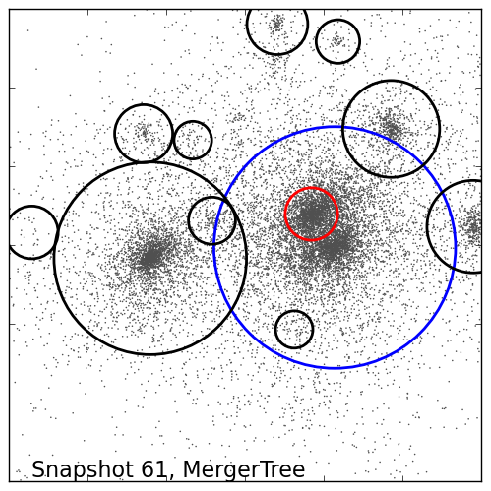}

  \vspace{2 mm}
  \includegraphics[width=1\figtable]{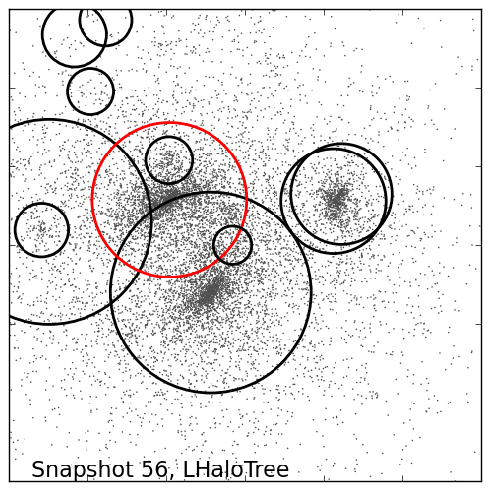}
  \includegraphics[width=1\figtable]{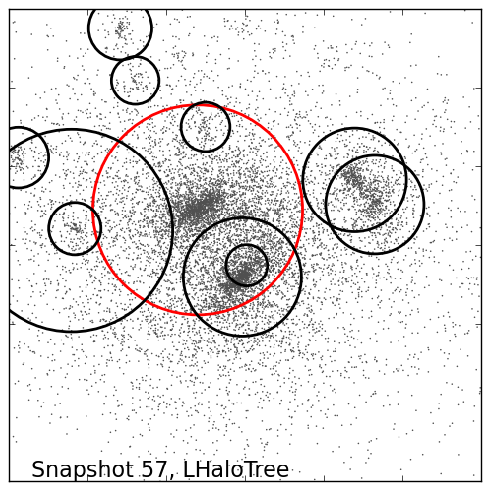}
  \includegraphics[width=1\figtable]{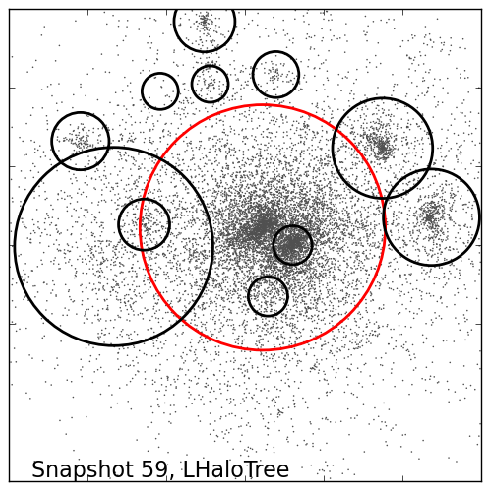}
  \includegraphics[width=1\figtable]{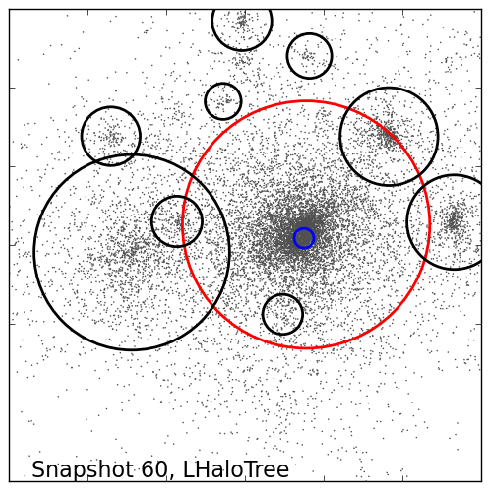}
  \includegraphics[width=1\figtable]{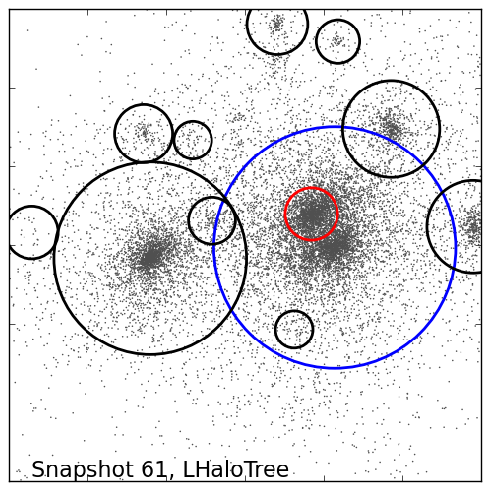}

  \vspace{2 mm}
  \includegraphics[width=1\figtable]{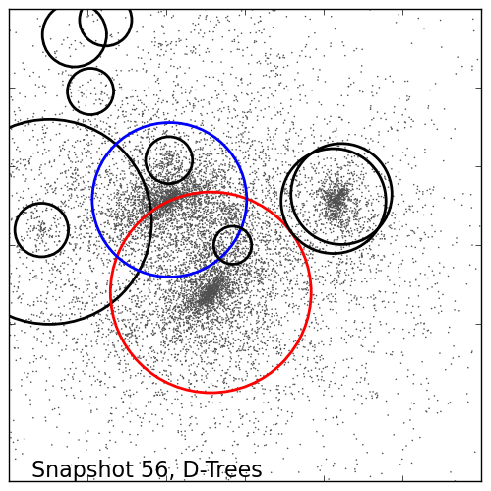}
  \includegraphics[width=1\figtable]{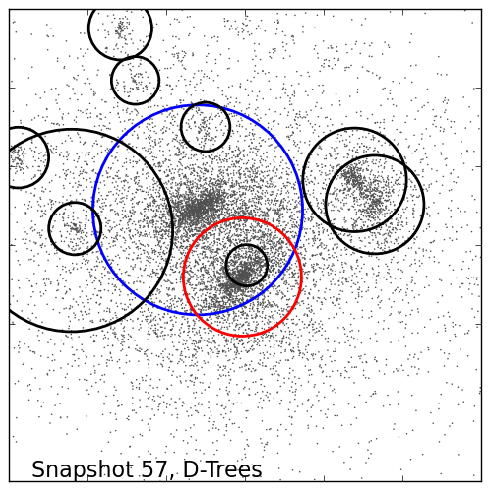} 
  \includegraphics[width=1\figtable]{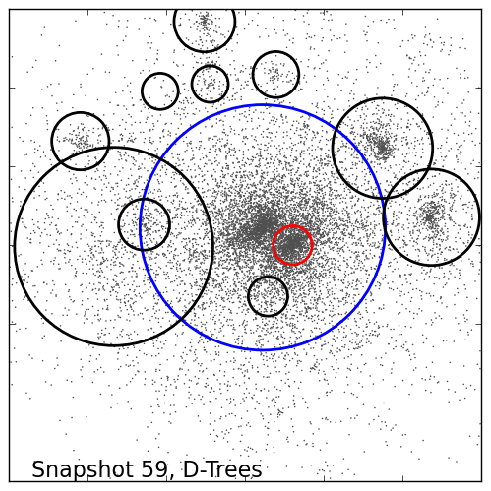}
  \includegraphics[width=1\figtable]{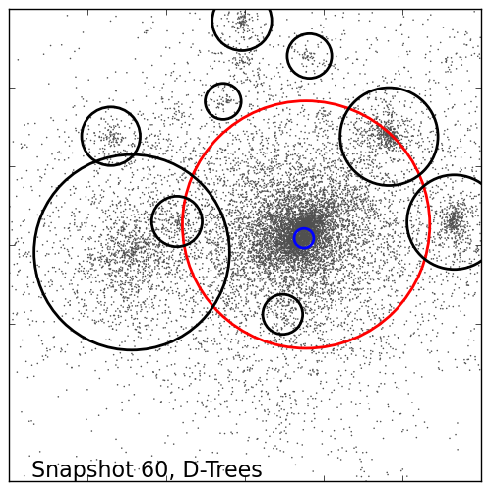} 
  \includegraphics[width=1\figtable]{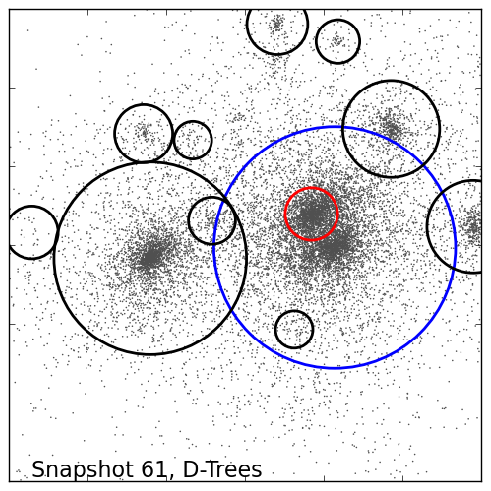}

  \vspace{2 mm}
  \includegraphics[width=1\figtable]{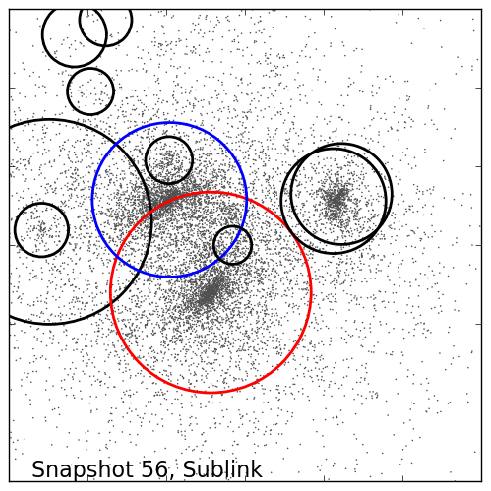}
  \includegraphics[width=1\figtable]{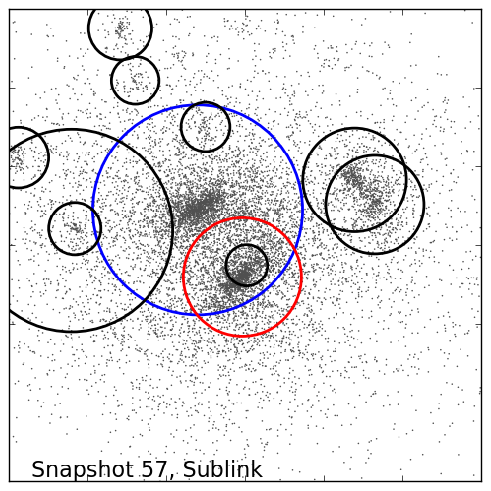} 
  \includegraphics[width=1\figtable]{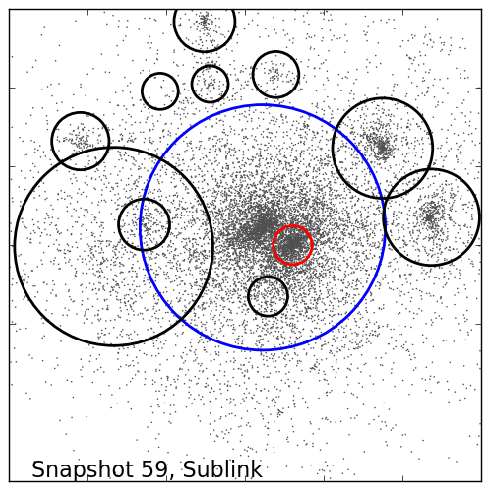}
  \includegraphics[width=1\figtable]{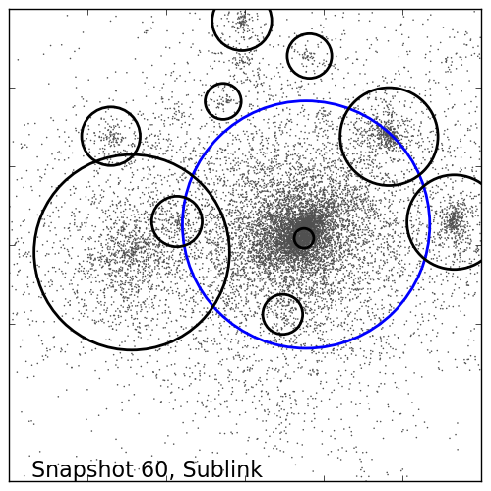}
  \includegraphics[width=1\figtable]{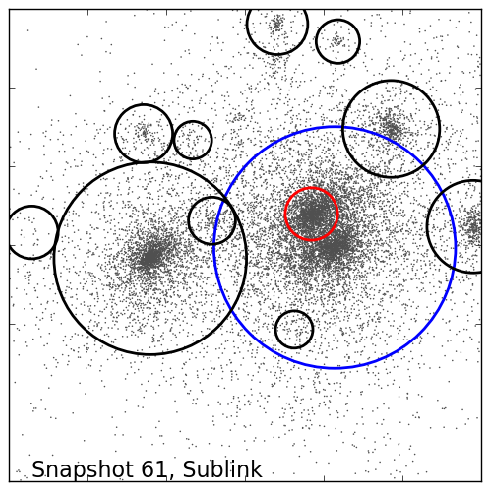}

  \vspace{2 mm}
  \includegraphics[width=1\figtable]{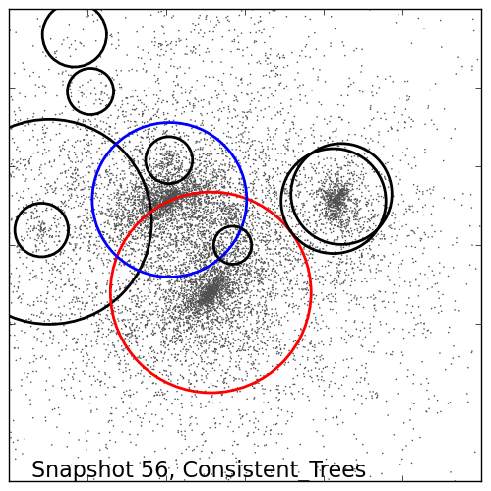}
  \includegraphics[width=1\figtable]{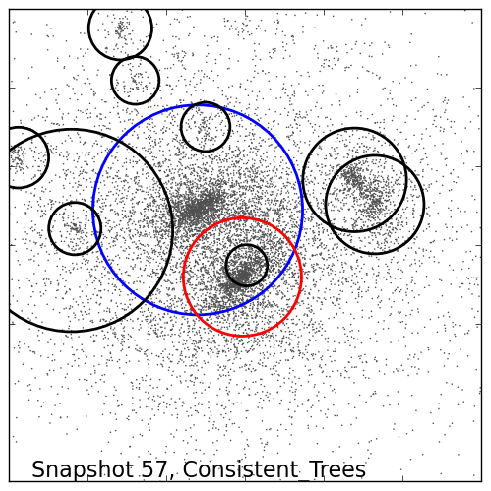} 
  \includegraphics[width=1\figtable]{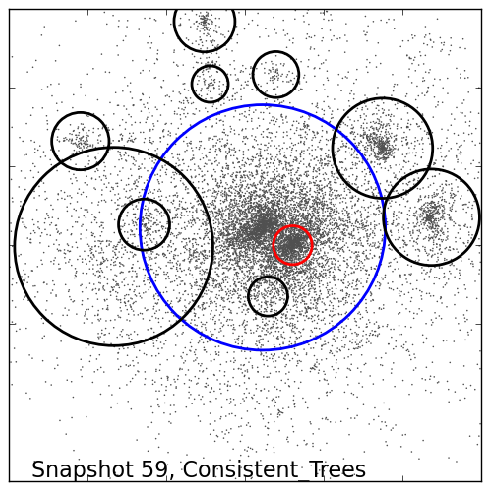}
  \includegraphics[width=1\figtable]{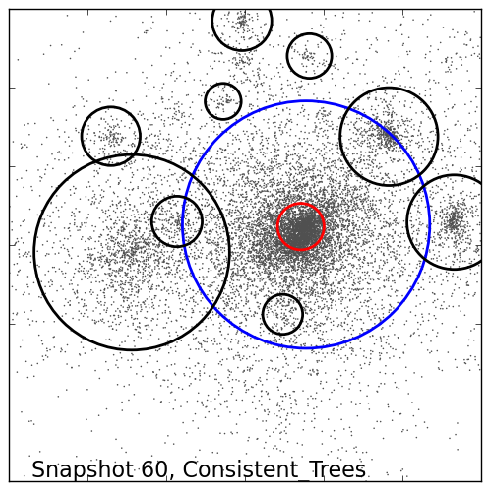} 
  \includegraphics[width=1\figtable]{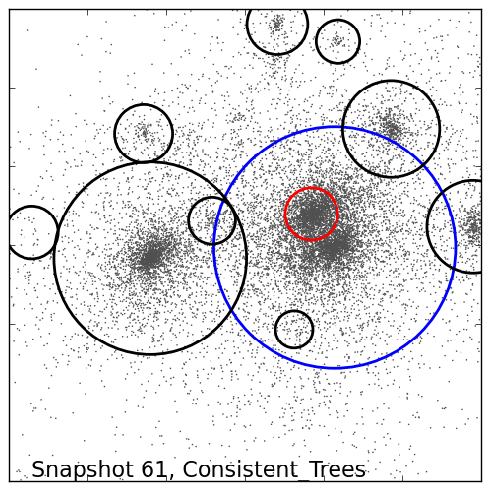}

  \vspace{2 mm}
  \includegraphics[width=1\figtable]{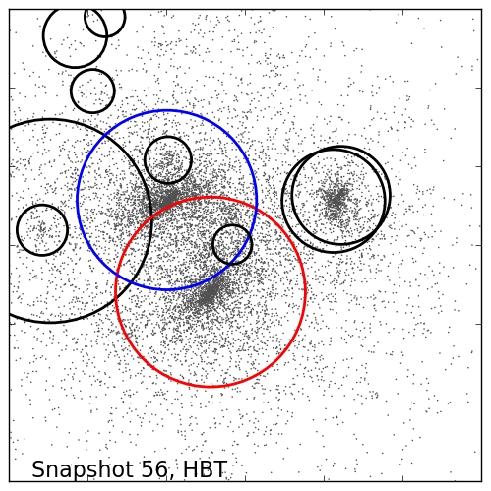}
  \includegraphics[width=1\figtable]{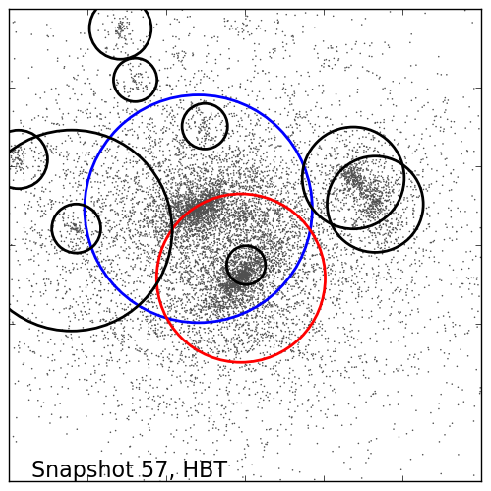} 
  \includegraphics[width=1\figtable]{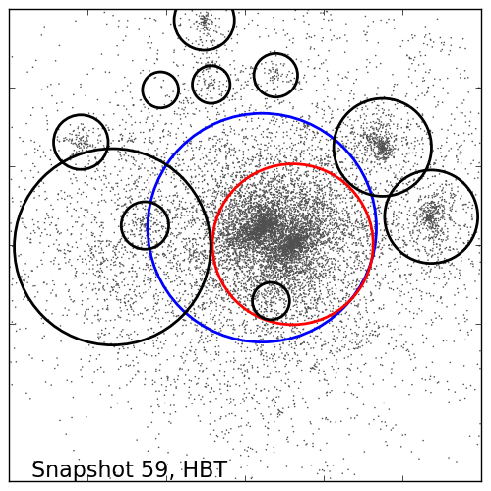}
  \includegraphics[width=1\figtable]{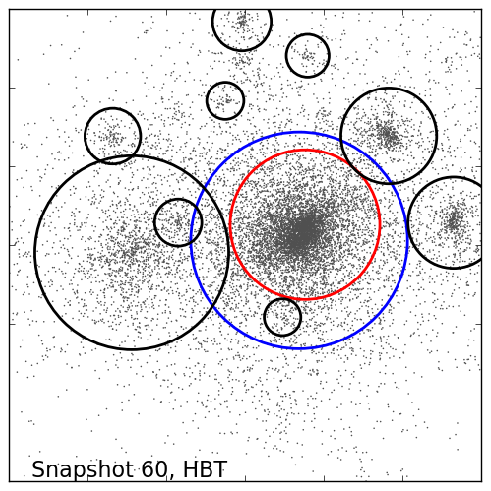} 
  \includegraphics[width=1\figtable]{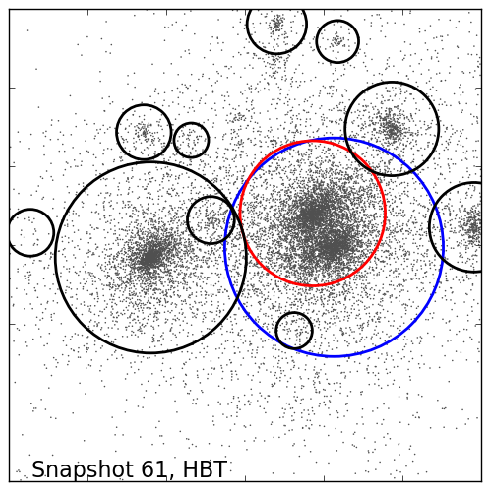}

  \caption{An example of the merger of two \halos\ where 
    \resubmit{the fluctuation of centering and size causes difficulties}
    for the merger-tree algorithms.  The red and blue circles show two
    \halos\ selected at $z=0$ (right-hand column) and then traced back in
    time over several snapshots (successive columns to the left - note
    that we have chosen to omit Snapshot 58 as it added little to the
  plot).  The missing algorithms all return the same results as
  \mergertree, shown in the top row.  See the main text for a commentary.}
  \label{fig:mergesample_panels}
\end{figure*}

One such example is pictured in \Fig{fig:mergesample_panels} which
shows two similarly-sized \halos\ merging almost head-on.  The red and
blue circles show the two \halos\ at $z=0$ (right-hand column) and
then traced back in time over several snapshots (successive columns to
the left - note that we have chose to omit Snapshot 58 as it added
little to the plot).  The \ahf\ halo finder (and other halo finders
behave in a similar manner) assigns most of the mass in overlapping
objects to a single object, treating the other as substructure.
Unfortunately, this assignment can change between snapshots so that
\halos\ centred on the same clump of highly-bound particles can
fluctuate wildly in size.  Different tree codes handle this in
different ways, illustrated in the different rows of
\Fig{fig:mergesample_panels}.  
\begin{itemize}
\item \mergertree\ fails to find a match for the smaller of the two
  \halos\ at Snapshot 60 and does not seek a match at earlier times.
  This halo therefore has no links in its merger tree and appears to
  be created intact in the final snapshot.  The other merit function
  codes that use just 2 snapshots (\treemaker, \velociraptor\ and
  \ysamtm) behave in the same manner, as, in this case, does \jmerger.
\item \LHaloTree\ does something similar, but due to its use of 
  weighted function, it matches the smaller
  of the two \halos\ at $z=0$ to the large one from the previous
  snapshot.  While \LHaloTree\ can cross-match \halos by skipping a
  snapshot, that is not applied here as a descendent halo exists.
\item \dtree\ does the same as \LHaloTree\ on Snapshot 60, but also
  manages to link together the larger of the two \halos\ between
  Snapshots 61 \& 59.  This results in a fluctuating mass
  for the both \halos, (low-high-low for red, high-low-high for blue).
\item \sublink\ also manages to cross-match the larger of the \halos\ between Snapshots
  61 \& 59 but chooses a different association for the halo in
  Snapshot~60, thus avoiding the large mass fluctuation.  It links the
  smaller of the two halos in Snapshot~61 directly to that in
  Snapshot~59, skipping over the intermediate snapshot.
\item \consistenttree\ goes one step further and introduces a fake
  halo in Snapshot~60 to avoid a link in the merger tree that extends
  over more than one snapshot.
\item Finally, \hbt\ redefines both \halos\ and outputs a smoother
  variation of mass over time.
\end{itemize}
From these descriptions, it may seem like the above is an ordered list
of improving performance, from top to bottom.  However, we stress that
this is true only for this particular merging event and that different
codes cope better in different situations.  The purpose here was more
to illustrate the variety of behaviours that are possible.

\subsection{Branching ratio} \label{sec:branches}
Another interesting statistical quantity is the number of branches
(i.e.~ the number of direct progenitors) at every node of the merger tree.
This will depend upon the spacing between snapshots, and so the
precise values are not important, but the differences between
algorithms are still of interest.

\begin{figure}
 \includegraphics[width=\figwidth]{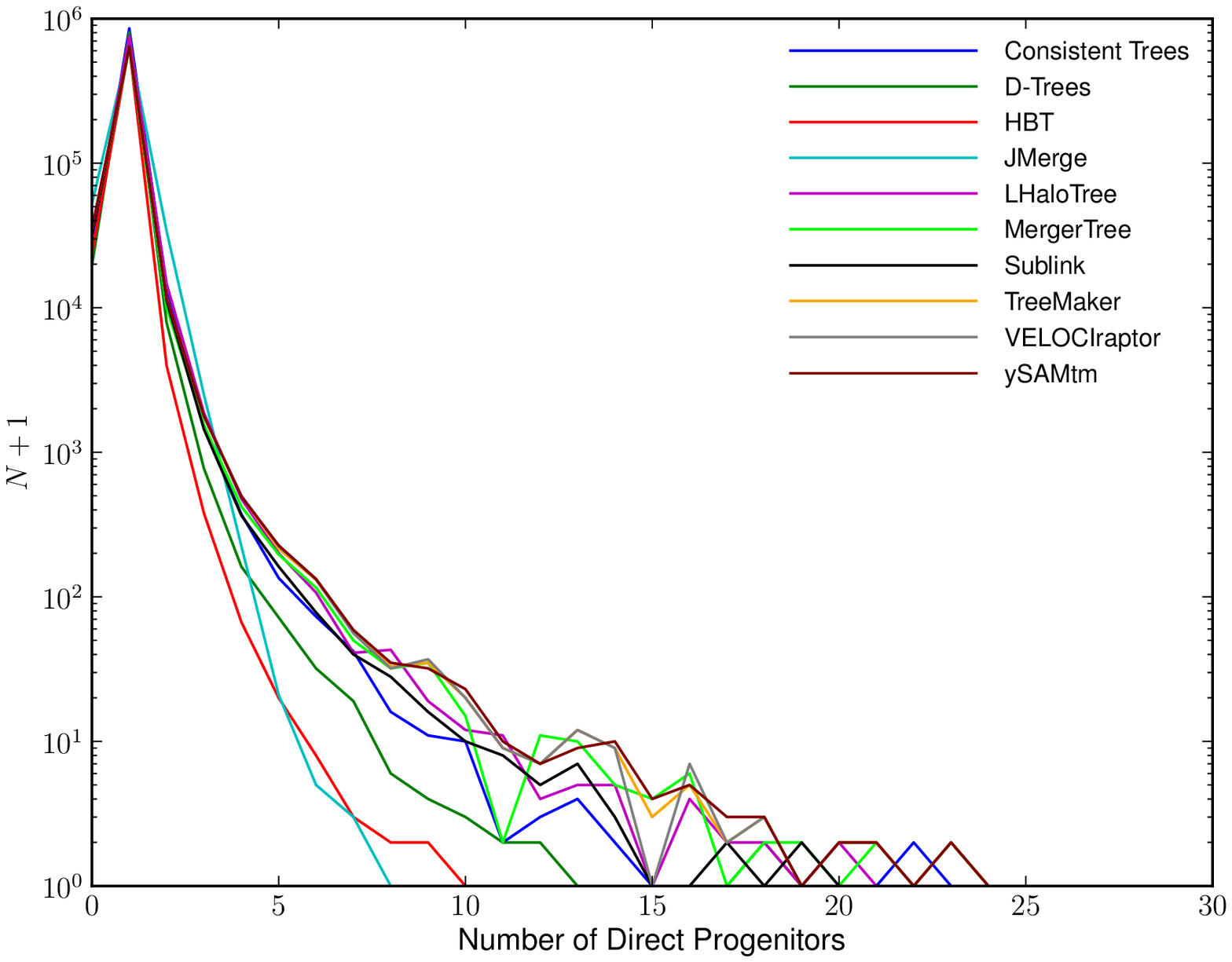}
\caption{Histograms of the number of \halos\ with $N_{\rm dprog}$
  direct progenitors, using all halos from $z=0$ to $z=2$.}
\label{fig:Nmergers}
\end{figure}

In \Fig{fig:Nmergers} we plot the number of tree nodes with $N_{\rm
dprog}$ direct progenitors, including all \halos\ between redshift
zero and two.  In this range the timestep $\Delta t$ between snapshots
is roughly constant with $\Delta t \sim 0.4$\,Gyr. The most common
situation is to have a single progenitor (i.e.~ the halo existed in
the previous snapshot but no merging took place), followed by zero
progenitors (i.e.~the halo appears for the first time).  However, in
some cases, and depending on the tree-builder, the number of direct
progenitors can exceed 20.

\hbt\ has the lowest branching ratio, perhaps because it allows itself
to modify the halo catalogue to extend the life of sub\halos.
\jmerger\ also has a low branching number because its non-use of
particle IDs gives it freedom to link together \halos\ that other
algorithms classify as unrelated.  Next come \dtree\ and
\consistenttree\ which both use information extended over several
timesteps to follow \halos\ that temporarily disappear (for instance
when a subhalo comes close to the centre of its host halo).

Although multiple direct progenitors are rare, it can be seen that the
choice of tree code can make a significant difference to the ability
to follow substructures and hence to the length of time a subhalo
exists before it is subsumed into the host halo.

\subsection{Misidentifications} \label{sec:misids}

Most tree-building algorithms link together \halos\ on the basis of
having particles in common. However, there are some that do not (in
this paper, \jmerger), and there are occasions when this association
is not clear-cut.  So we wish to test how often an obvious
mis-identification occurs.

One way of doing this is to quantify how far \halos\ are displaced from
their expected locations in moving from one snapshot to the next.
This is hard to predict for sub-\halos\ that may be moving around inside
a larger object and so we restrict our attention to main \halos\ only.
To measure this deviation we use the statistic
\begin{equation}
  \Delta_r = \frac{|\mathbf{r}_B-\mathbf{r}_A
    -0.5(\mathbf{v}_A+\mathbf{v}_B)\,(t_B-t_A)|}
    {0.5\,(R_{\mathrm{200}A}+R_{\mathrm{200}B} + |\mathbf{v_A+
        \mathbf{v_B}}|(t_B - t_A))}
\end{equation}
which stays small as long as there is approximately uniform
acceleration and no error in the halo linking.  Here $t$ is cosmic
time, $\mathbf{r}$ \& $\mathbf{v}$ are the \halos' positions and
velocities, and $R_\mathrm{200}$ the radius that encloses an
overdensity of 200 times the critical density.  The subscripts $A$ and
$B$ refer to two linked \halos\ along the main branch of any tree.

\begin{figure}
 \includegraphics[width=\figwidth]{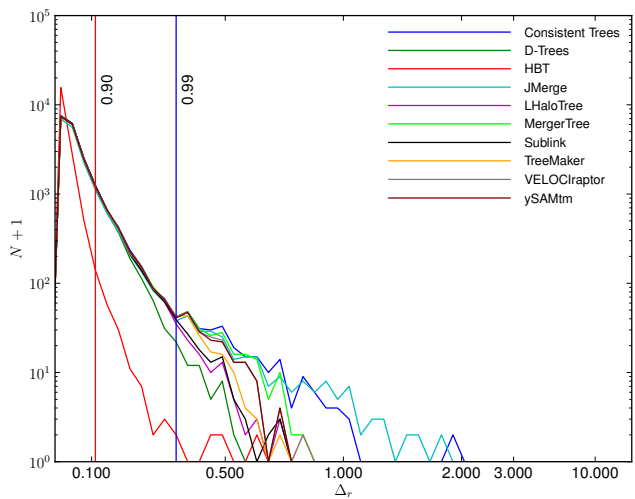}
\caption{Histograms of the displacement statistic, $\Delta_r$, for
main \halos\ and their main progenitor for which both of them have
$M_{\rm{200}} > 10^{12}\,h^{-1}M_\odot$.  The vertical lines show the
  90$^\mathrm{th}$ and 99$^\mathrm{th}$ percentiles for
  \mergertree\ (but are approximately the same for all
  algorithms except \hbt).}
\label{fig:disp_main_host}
\end{figure}

\Fig{fig:disp_main_host} shows a histogram of $\Delta_r$ for each
algorithm, for all main \halos\ and their corresponding main
progenitors.  Most algorithms agree on the bulk of the distribution,
and this likely represents the true behaviour for the
\ahf\ \halos\ considered here,
with deviations from $\Delta_r=0$ being caused by curved trajectories
and/or merging of sub\halos.  The difference in \hbt's result from the
others is partly due to different tree-links but also because the
\hbt\ halo catalogue has an intrinsically lower $\Delta_r$.

\begin{figure}
  \centering
  \vspace{2 mm}
  \includegraphics[width=0.85\figtable]{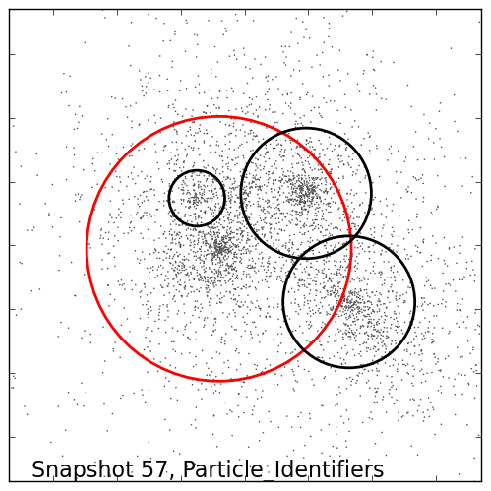}
  \includegraphics[width=0.85\figtable]{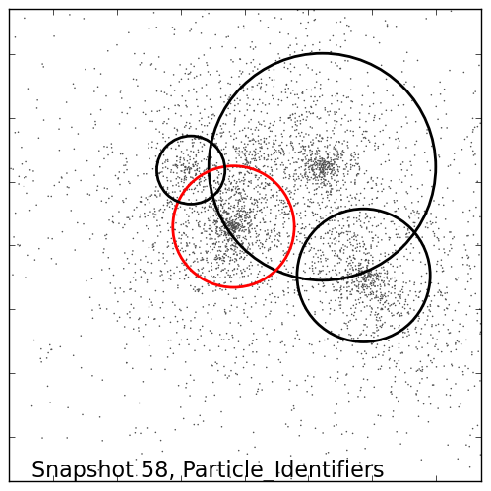}
  \includegraphics[width=0.85\figtable]{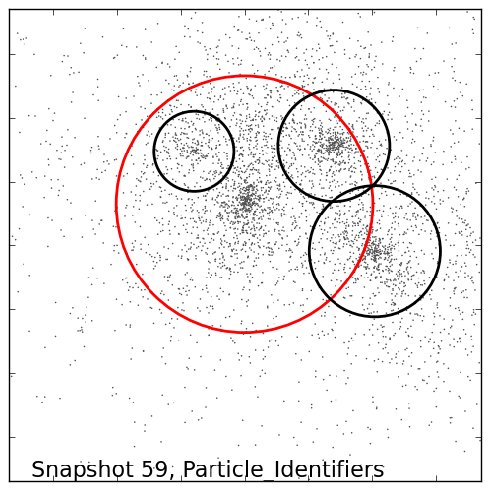}

  \vspace{2 mm}
  \includegraphics[width=0.85\figtable]{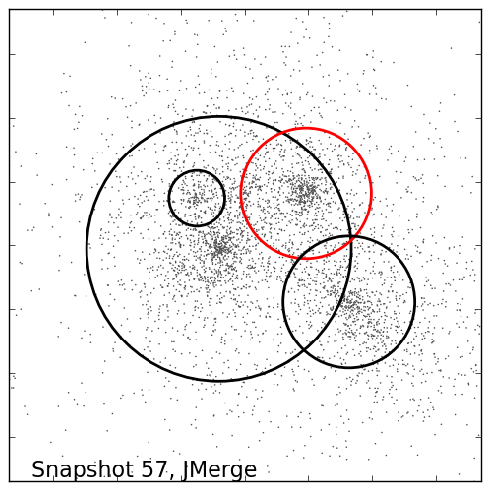}
  \includegraphics[width=0.85\figtable]{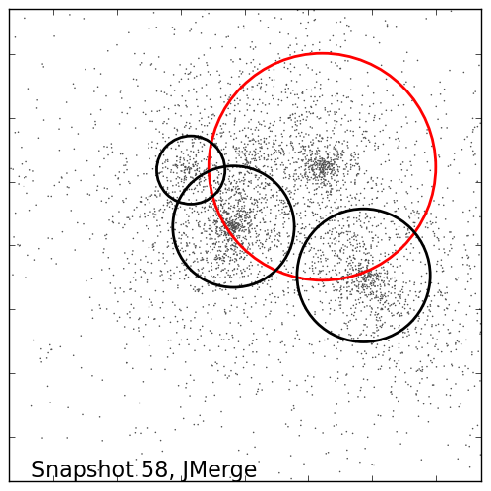} 
  \includegraphics[width=0.85\figtable]{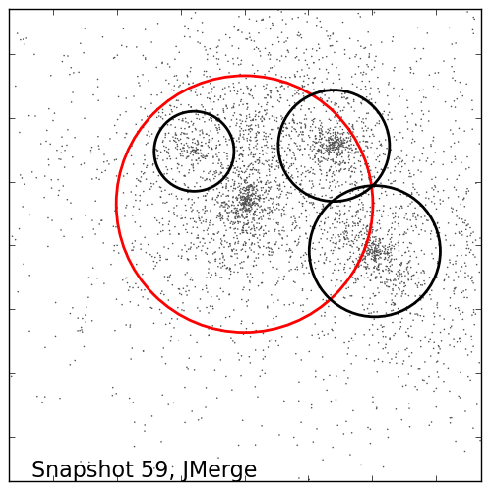}

  \vspace{2 mm}
  \includegraphics[width=0.85\figtable]{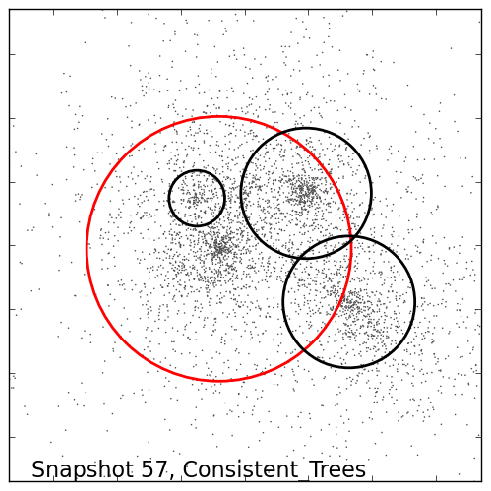}
  \includegraphics[width=0.85\figtable]{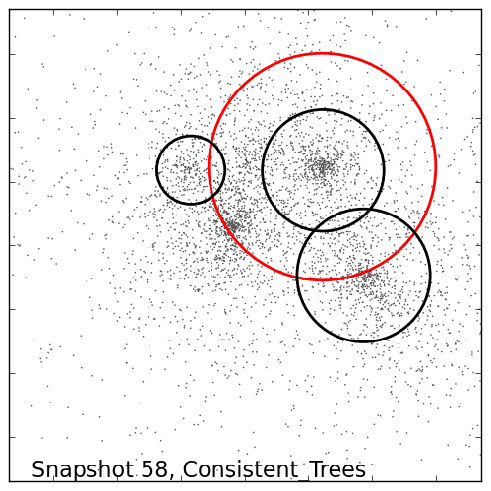} 
  \includegraphics[width=0.85\figtable]{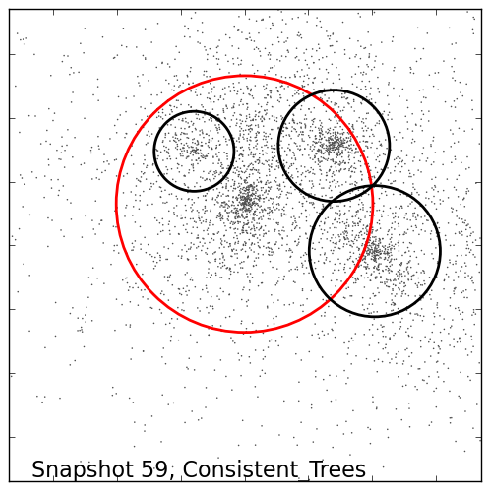}

  \vspace{2 mm}
  \includegraphics[width=0.85\figtable]{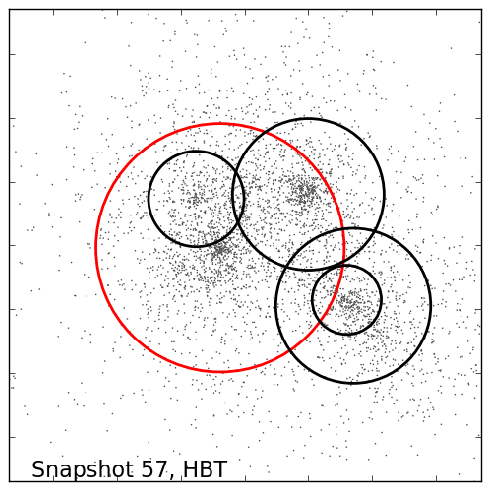}
  \includegraphics[width=0.85\figtable]{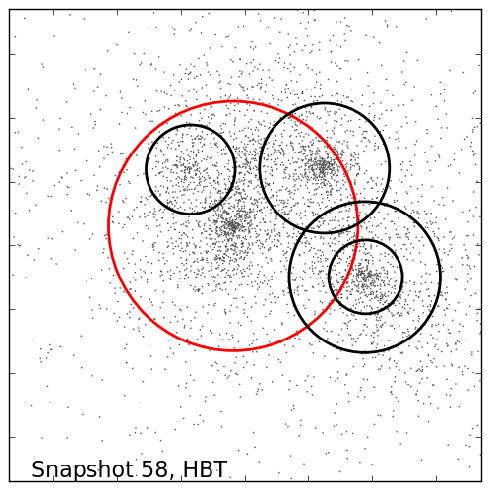} 
  \includegraphics[width=0.85\figtable]{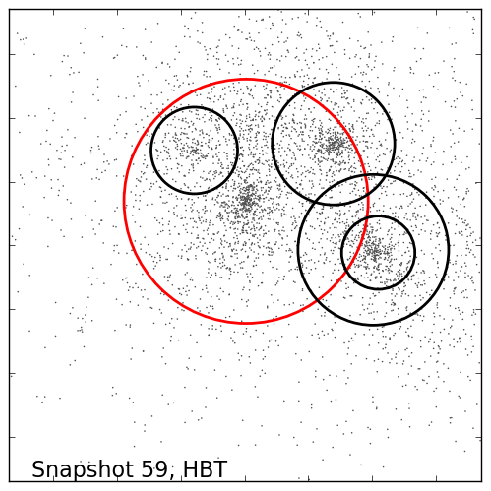}
  \caption{An example of a situation where the halo finder
    assigns main halos differently between snapshots.  The red
    \halos\ in each row show the main branch of the largest halo on
    the right-hand side.}
\label{fig:dumbellsample_panels}
\end{figure}

\jmerger\ occasionally shows much larger deviations, suggesting that
it does have a tendency to link together unassociated \halos.  
\consistenttree\ also shows large outliers in this test and
\Fig{fig:dumbellsample_panels} shows a typical example of how this
comes about.  Here we see an interaction in which the assignment of
main halo alternates between successive snapshots:
\begin{itemize}
\item Most algorithms (top row) link together the visually correct
  group of particles and have small $\Delta_r$, but will have a large
fluctuation in halo mass along the main branch.
\item \jmerger\ requires smooth changes in mass and so it follows the
  main halo between Snapshots~58 \& 59, leading to a large value of
  $\Delta_r$.
\item \consistenttree\ follows the main branch across all three
  snapshots, giving large values of $\Delta_r$ for both links.
It (correctly) fails to associate the top-right halo in Snapshot~59
with the central one in Snapshot~58, so it removes the latter and
creates a fake halo to take its place.
\item \hbt\ resolves the situation by creating a halo catalogue in
  which the mass evolution is smoother.  It also inserts an
  extra subhalo on the bottom-right that is not returned by any of
  the other algorithms. 
\end{itemize}

\subsection{The loss of particles during halo growth}

During mergers (and, indeed, during quiescent evolution) particles can
be lost from \halos.  As a measure of this, we use the statistic
\begin{equation}
 \Delta_N = \frac{N_{\cup A_i} - N_{(\cup A_i)\cap B}}{N_{\cup A_i}} ,
\end{equation}
where, for a given halo $B$, the union runs over all direct progenitors,
$A_i$.  Here $N$ is the number of particles in $\cup A_i$ and $B$ or common to
them both, as indicated by the subscript.

\begin{figure}
 \includegraphics[width=\figwidth]{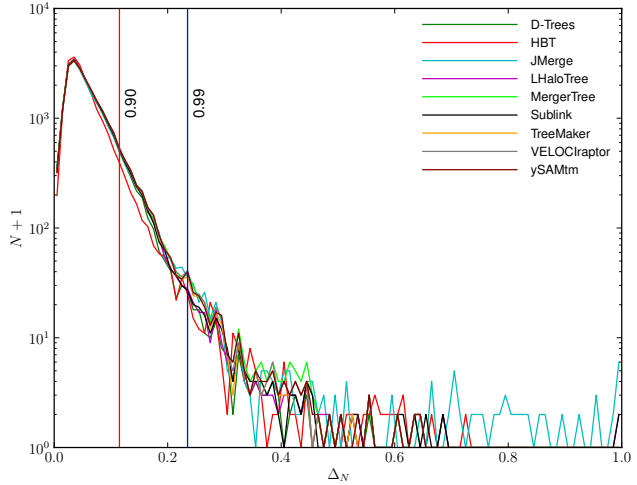}
\caption{The distribution function of the fraction of lost particles,
  $\Delta_N$ for \halos\ along the main branch with $M_{\rm{200}} >
  10^{12}\,h^{-1}M_\odot$. The vertical lines show the
  90$^\mathrm{th}$ and 99$^\mathrm{th}$ percentiles for
  \mergertree\ (but are approximately the same for all
  algorithms). Please note that \consistenttree\ cannot be included in
  this test because the added halos specified by the code do not have
  particle information.}
\label{fig:lostmass_host_new}
\end{figure}

The distribution function of the fraction of lost particles,
$\Delta_N$ for \halos\ along the main branch with $M_{\rm{200}} >
10^{12}\,h^{-1}$\Msun\ (corresponding to about 1000 particles) is
shown in \Fig{fig:lostmass_host_new}.  Note the extensive wing on this
plot that extends to $\Delta_N=0.4$.  For small values of $\Delta_N$,
this is due to changes in the shape of the halo, and to natural
particle orbits that results in material moving out across the radius
(here $R_{200}$) used to define the edge of the halo.  Large values of
$\Delta_N$ can occur when \halos\ reduce their size significantly
between snapshots.  An example of this situation has already been
shown in the third row of \Fig{fig:mergesample_panels} which
illustrates how the halo finder alternates between allocating most of
the mass to one or other of two \halos\ as they fly by one another.

All halo finders roughly agree on the number of \halos\ for which
$\Delta_N<0.4$, but there are signficant differences for larger
values -- these are most probably due to mis-identifications.  It is
perhaps not surprising that \jmerger\ has occasional very poor
matches, given that it does not use particle IDs, but rare examples of
apparently erroneous links are found in many other algorithms too.

\section{Mass growth}\label{sec:mass}

In this section we look at the mass evolution of \halos, primarily
along their main branches, which is a key input for most \SAMs.  While
main \halos\ are expected to grow in mass through accretion and
mergers, sub-\halos\ can lose mass through tidal stripping.

\begin{figure}
  \centering
  \includegraphics[width=1\linewidth]{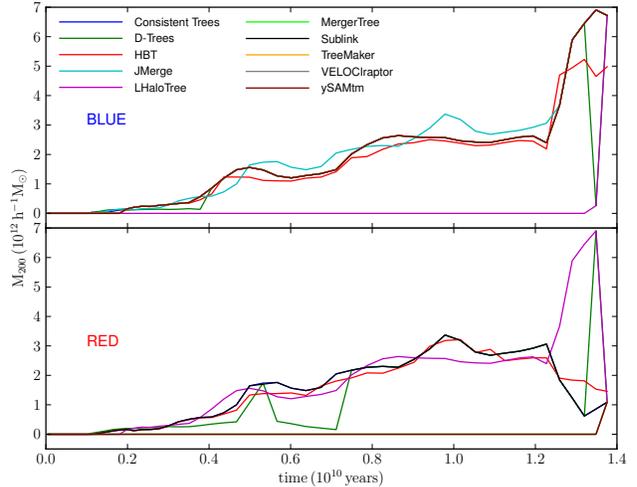}
  \caption{The mass history of the blue halo (top) and the red halo
    (bottom) in \Fig{fig:mergesample_panels} specified by each
    merger-tree code.  Note that many of the lines lie on top of one
    another - we do not attempt to describe that in detail here as the
  purpose of the plot is simply to illustrate the variety of
  mass-accretion histories that are possible for a single halo.  The
  \hbt\ \halos\ end up with a different final mass at $z=0$ because
  they produce a distinct halo catalogue.}
  \label{fig:mbmass}
\end{figure}

Consider first \Fig{fig:mbmass} which shows the mass evolution along
the main branch for the red and blue \halos\ illustrated
\Fig{fig:mergesample_panels}.  The large mass fluctuations seen on the
right-hand side of this plot correspond to the rightmost panels in
\Fig{fig:mergesample_panels} and illustrate how poorly-constrained the
mass evolution is during that merger.  \resubmit{The strong variation
  between the results returned by different algorithms suggests that
  much of this mass variation is unphysical, and} most \SAMs\ would
struggle to cope with this kind of fluctuating mass behaviour.

\subsection{Mass growth along the halo main branch}\label{sec:massmain}

\begin{figure}
 \includegraphics[width=\figwidth]{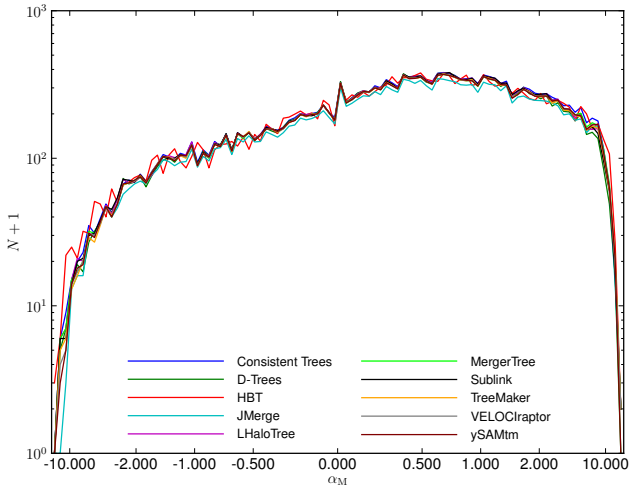}
\caption{Distribution function of logarithmic mass growth,
  $\alpha_\mathrm{M}$ along halo main branches.  We have
  included all pairs of \halos\ for which both the masses exceed
  $10^{12}$ $h^{-1}M_\odot$.}
\label{fig:alpham}
\end{figure}

The logarithmic growth rate of main branch \halos, d$\log M$/d$\log t$ is
approximated discretely by
\begin{equation}
 {\mathrm{d}\log M\over\mathrm{d}\log t} \approx
 \alpha_\mathrm{M}(A,B)=
 \frac{(t_B+t_A)(M_B-M_A)}{(t_B-t_A)(M_B+M_A)},
\label{eq:deltam}
\end{equation}
where $M_A$ and $M_B$ are the masses of a halo and its descendent at
times $t_A$ and $t_B$, respectively.  The distribution function of
$\alpha_{\rm M}$ is shown in \Fig{fig:alpham} for every pair
of main-branch \halos\ for which the mass of each exceeds
$10^{12}\, h^{-1}\mathrm{M}_\odot$ (corresponding to about 1000 particles).

As demonstrated in \Fig{fig:alpham}, most of the time \halos\ are
growing but there is a significant proportion of the time (about 30\%)
during which mass loss occurs.  Such a large fraction is unlikely to
be due to stripping (as this result is restricted to high-mass
main-branch \halos) but some apparent mass loss can occur due to
changes in the shape of \halos\ during their evolution, especially
following a major merger.

Strong mass loss, however, is unphysical and is due to failures in the
halo-finding and linking process, as illustrated in
\Figs{fig:mergesample_panels}, \ref{fig:dumbellsample_panels} \&
\ref{fig:mbmass}.  The halo evolution seen in the rightmost columns of
\Fig{fig:mergesample_panels} correspond to the wings in
\Fig{fig:alpham}.

\subsection{Mass fluctuations of subhalo main branches}\label{sec:massfluct}

Abrupt fluctuations up and down in mass can be quantified
with a statistic
\begin{equation}
  \xi_M(k) = \arctan\alpha_M(k,k+1) - \arctan\alpha_M(k-1,k).
\end{equation}
where $\alpha_M$ is as defined in \Eq{eq:deltam} and $k-1$, $k$ \&
$k+1$ represent successive timesteps.  This measures the change
in the slope of the mass accretion rate between two
consecutive steps and thus ranges from $-\pi$ to $\pi$.  The main
purpose of this statistic is to detect temporary mass fluctuations
that occur either as a result of the natural growth process, or because
of halo misidentification.

\begin{figure}
 \includegraphics[width=\figwidth]{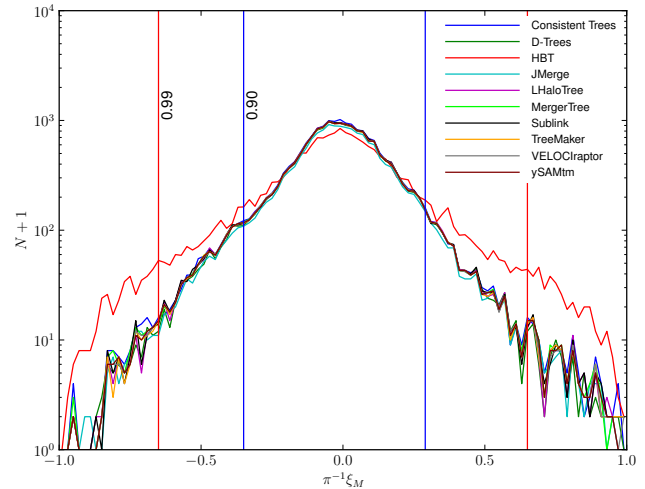}
\caption{Mass fluctuations, $\xi_M$, for sets of 3 consecutive
  \halos\ along a main branch for which the mass of each exceeds
  $10^{12}$ $h^{-1}M_\odot$.  The vertical lines show the two-sided
  90$^\mathrm{th}$ and 99$^\mathrm{th}$ percentiles for
  \mergertree\ (but are approximately the same for all
  algorithms except \hbt). 
  Note that the apparent discrepancy of
  \hbt\ is because, for the purposes of this paper, they construct
  masses only from the supplied \ahf\ halo catalogues.  We have
  checked that, on applying \hbt\ to the full simulation data, this
  discrepancy goes away.
}
\label{fig:mass_fluct}
\end{figure}

Large, negative values of $\xi_M$ correspond to sharp temporary peaks
in mass, and positive values to dips in mass.  Somewhat surprisingly
$|\xi_M|$ exceeds $\pi/3$ 10 per cent of the time, and $2\pi/3$ 1 per
cent of the time. Thus strong mass variations are relatively common.
\resubmit{However, the presence of the strong mass variations seems to be a 
limitation of the halo finding algorithm rather than the merger tree 
algorithms as evidenced by the great similarity
between all merger tree algorithms except \hbt\ in \Fig{fig:mass_fluct}.}
Note that the apparent discrepancy of \hbt\ is because, for the
purposes of this paper, they construct masses only from the supplied
\ahf\ halo catalogues.  We have checked that, on applying \hbt\ to the
full simulation data, this discrepancy goes away.

\section{Discussion} \label{sec:disc}

This paper summarises the results of a merger tree comparison
project. The comparison was completed, and the paper drafted, in
advance of the {\sc Sussing Merger Trees} Workshop in Midhurst, Sussex
in July 2013. The aim of the workshop was not only to compare the
existing status of merger tree codes, but also to get people thinking
about the desirable features of such codes, in particular for their use
as backbones for \SAM{}ling.

Ten different merger tree builders contributed to this comparison
project, as listed in \Tab{tbl:Requirements}.  Although many of these
adopted similar approaches, no two gave identical results.

In order to enable the comparison, we desired that each merger tree
code should use the same \halos\ as input.  It soon became apparent
that the halo finder can be intimately linked to the tree-builder
itself, and so some tree-building codes needed modification to enable
them to take part.  For two of the codes (\consistenttree\ \& \hbt),
we had to allow modification of the halo catalogue.  For this reason,
and because the quality of a merger tree depends in some unspecified
way upon the particular scientific use to which it will be put, we
avoid making conclusive statements here about which algorithms perform
better than others.

In \Sec{sec:terminology}, we defined some terminology that we used
throughout the paper.  This proved essential to get everyone talking a
common language (for example, some algorithms did not initially return
merger \textit{trees} at all, in the sense that every halo did not
have a unique descendent).  We encourage other members of the
community to use the same nomenclature.

\subsection{Summary of results} \label{sec:summary}

Here we present a brief summary of our findings:
\begin{itemize}
\item Imperfections in the halo finder can lead to great difficulties
  for tree-building algorithms.  The particular halo finder that we
  used in this project was \ahf, but we would expect similar behaviour
  with other halo finders and a study of this is presently under way.
\item The temporary loss of a halo during the merger of two \halos\
  (see, e.g.~\Fig{fig:mergesample_panels}) is disastrous for
  tree-building algorithms that examine only two adjacent snapshots.
  In such cases, it is possible for \halos\ containing over 1000
  particles to apparently appear out of nothing between two adjacent
  snapshots.
\item Although they were working with the same input halo catalogue,
  different algorithms varied in their ability to link together
  sub\halos, leading to significantly different branching ratios for
  the trees.
\item Due to the limitations of the halo finder, codes that do not use
  particle IDs to link together \halos\ can occasionally produce clear
  mis-identifications (see, e.g.~\Fig{fig:dumbellsample_panels}).
\item Even when \halos\ persist between snapshots, the halo finder will
  sometimes alter which of the two it treats as the main halo, and
  this can lead to large oscillations in mass.  Different
  tree-builders handle this in different ways.
\item The slope of the logarithmic mass growth curve, d$\log M$/d$\log
  t$ has a very broad distribution with a peak around 0.5 to 1 but
  extending beyond the range $-10$ to 10.  Much of this is due to
  genuine fluctuations in mass, although the extremes are due to
  failures in the combined halo finder and tree builder.
\end{itemize}
We suggest that any optimal tree-building algorithm will require a
high-quality input halo catalogue that minimises 'lost' \halos\ and mass
fluctutations, and in addition will possess the following:
\begin{itemize}
\item the use of particle IDs to match \halos\ between snapshots;
\item the ability to skip at least one, and preferably more, snapshots
  in order to recover sub\halos\ that are temporarily lost by the
  halo finder (for instance when they transit the centre of the host halo);
\item the ability to cope with (and ideally smooth out) large,
  temporary flucuations in halo mass.
\end{itemize}

\subsection{Future work} \label{sec:future}

One of the main purposes of the workshop was to stimulate people into
thinking harder about what makes a good merger tree.  As a result of
this, we have initiated projects on the following topics:
\begin{itemize}
\item Tree stability versus number of snapshots and their optimal
  spacing.
\item Which is the best halo finder to use for the purposes of
  tree building?  The answer to this question may well vary from one
  tree-building code to another.
\item Related to the above, what is the best overdensity criterion to
  use when defining \halos?
\item How do the results change when applied to a large resimulation
  of a single halo with lots of nested substructure?
\item What is the effect of the variation in merger trees on the
  resultant \SAMs?
\end{itemize}

It is our hope that a consensus will emerge, if not on a unique halo
finding and merger tree algorithm, at least upon the desirable
features that such algorithms should possess in order to obtain stable
results for the purposes of \SAM{}ling.

\section*{Acknowledgements} \label{sec:Acknowledgements}

The {\sc Sussing Merger Trees} Workshop was supported by the European
Commission's Framework Programme 7, through the Marie Curie Initial
Training Network CosmoComp (PITN-GA-2009-238356).  This also provided
fellowship support for AS.

PSB received support from HST Theory Grant HST-AR-12159.01-A,
provided by NASA through a grant from the Space Telescope Science
Institute, which is operated by the Association of Universities for
Research in Astronomy, Incorporated, under NASA contract NAS5-26555.

KD acknowledges the support by the DFG Cluster of Excellence 'Origin
and Structure of the Universe'. 

PJE is supported by the SSimPL programme and the Sydney Institute for
Astronomy (SIfA).  

JXH is supported by an STFC Rolling Grant to the Institute for
Computational Cosmology, Durham University.

YPJ is sponsored by NSFC (11121062 11033006) and the CAS/SAFEA
International Partnership Program for Creative Research Teams
(KJCX2-YW-T23).

AK is supported by the {\it Spanish Ministerio de Ciencia e
  Innovaci\'on} (MICINN) in Spain through the Ram\'{o}n y Cajal
programme as well as the grants AYA~2009-13875-C03-02, CSD2009-00064,
CAM~S2009/ESP-1496 (from the ASTROMADRID network) and the {\it
  Ministerio de Econom\'ia y Competitividad} (MINECO) through grant
AYA2012-31101. He further thanks Curtis Mayfield for superfly.

VRG was supported in part by Consejo Nacional de Ciencia y
Tecnolog\'ia (CONACyT) and Fundaci\'on M\'exico en Harvard.

CS is supported by the Development and Promotion of Science and Technology 
Talents Project (DPST), Thailand.

PAT acknowledges support from the Science and Technology Facilities
Council (grant number ST/I000976/1).

SKY acknowledges support from National Research Foundation of Korea
(Doyak Program No.~20090078756; SRC Program No.~2010-0027910) and DRC
Grant of Korea Research Council of Fundamental Science and Technology
(FY~2012). Numerical simulation was performed using the KISTI
supercomputer under the program of KSC-2012-C2-11 and
KSC-2012-C3-10. Much of this manuscript was written during the visit
of SKY to the Universities of Nottingham and Oxford under the
general support of the LG Yon-Am Foundation. 

YYM received support from the Weiland Family Stanford Graduate Fellowship.

The authors contributed in the following ways to this paper: CS, AK,
FRP, AS, PAT organised this project. They designed the comparison,
planned and organised the data, performed the analysis presented and
wrote the paper. CS is a PhD student supervised by PAT.  The other
authors (as listed in \Sec{sec:tree}) provided results and
descriptions of their respective algorithms; they also helped to 
proof-read the paper.


\bibliography{mn-jour,treebuild}

\begin{thebibliography}{24}
\expandafter\ifx\csname natexlab\endcsname\relax\def\natexlab#1{#1}\fi

\bibitem[{{Aubert} {et~al.}(2004){Aubert}, {Pichon}, \& {Colombi}}]{Aubert2004}
{Aubert} D., {Pichon} C., {Colombi} S., 2004, MNRAS, 352, 376

\bibitem[{{Baugh}(2006)}]{Bau06}
{Baugh} C.~M., 2006, Reports on Progress in Physics, 69, 3101

\bibitem[{{Behroozi} {et~al.}(2013){Behroozi}, {Wechsler}, {Wu}, {Busha},
  {Klypin}, \& {Primack}}]{Behroozi_etal_2013}
{Behroozi} P.~S., {Wechsler} R.~H., {Wu} H.-Y., {Busha} M.~T., {Klypin} A.~A.,
  {Primack} J.~R., 2013, ApJ, 763, 18

\bibitem[{{Davis} {et~al.}(1985){Davis}, {Efstathiou}, {Frenk}, \&
  {White}}]{Davis1985}
{Davis} M., {Efstathiou} G., {Frenk} C.~S., {White} S.~D.~M., 1985, ApJ, 292,
  371

\bibitem[{{Elahi} {et~al.}(2013){Elahi}, {Han}, {Lux}, {Ascasibar}, {Behroozi},
  {Knebe}, {Muldrew}, {Onions}, \& {Pearce}}]{Elahi_2013}
{Elahi} P.~J., {Han} J., {Lux} H., {Ascasibar} Y., {Behroozi} P., {Knebe} A.,
  {Muldrew} S.~I., {Onions} J., {Pearce} F., 2013, submitted to MNRAS,
  arXiv:1305.2448

\bibitem[{Gill {et~al.}(2004)Gill, Knebe, \& Gibson}]{gill_evolution_2004}
Gill S.~P., Knebe A., Gibson B.~K., 2004, MNRAS, 351, 399

\bibitem[{{Han} {et~al.}(2012){Han}, {Jing}, {Wang}, \& {Wang}}]{han_etal_2012}
{Han} J., {Jing} Y.~P., {Wang} H., {Wang} W., 2012, MNRAS, 427, 2437

\bibitem[{{Hatton} {et~al.}(2003){Hatton}, {Devriendt}, {Ninin}, {Bouchet},
  {Guiderdoni}, \& {Vibert}}]{Hatton2003}
{Hatton} S., {Devriendt} J.~E.~G., {Ninin} S., {Bouchet} F.~R., {Guiderdoni}
  B., {Vibert} D., 2003, MNRAS, 343, 75

\bibitem[{{Klimentowski} {et~al.}(2010){Klimentowski}, {{\L}okas}, {Knebe},
  {Gottl{\"o}ber}, {Martinez-Vaquero}, {Yepes}, \&
  {Hoffman}}]{Klimentowski_etal_2010}
{Klimentowski} J., {{\L}okas} E.~L., {Knebe} A., {Gottl{\"o}ber} S.,
  {Martinez-Vaquero} L.~A., {Yepes} G., {Hoffman} Y., 2010, MNRAS, 402, 1899

\bibitem[{Knebe {et~al.}(2011)Knebe, Knollmann, Muldrew, Pearce,
  {Aragon-Calvo}, Ascasibar, Behroozi, Ceverino, Colombi, \&
  Diemand}]{knebe_haloes_2011}
Knebe A., Knollmann S.~R., Muldrew S.~I., Pearce F.~R., {Aragon-Calvo} M.~A.,
  Ascasibar Y., Behroozi P.~S., Ceverino D., Colombi S., Diemand J., 2011,
  MNRAS, 415, 2293

\bibitem[{{Knebe} {et~al.}(2013{\natexlab{a}}){Knebe}, {Libeskind}, {Pearce},
  {Behroozi}, {Casado}, {Dolag}, {Dominguez-Tenreiro}, {Elahi}, {Lux},
  {Muldrew}, \& {Onions}}]{Knebe_galaxies_2013}
{Knebe} A., {Libeskind} N.~I., {Pearce} F., {Behroozi} P., {Casado} J., {Dolag}
  K., {Dominguez-Tenreiro} R., {Elahi} P., {Lux} H., {Muldrew} S.~I., {Onions}
  J., 2013{\natexlab{a}}, MNRAS, 428, 2039

\bibitem[{{Knebe} {et~al.}(2013{\natexlab{b}}){Knebe}, {Pearce}, {Lux},
  {Ascasibar}, {Behroozi}, {Casado}, {Corbett Moran}, {Diemand}, {Dolag},
  {Dominguez-Tenreiro}, {Elahi}, {Falck}, {Gottloeber}, {Han}, {Klypin},
  {Lukic}, {Maciejewski}, {McBride}, {Merchan}, {Muldrew}, {Neyrinck},
  {Onions}, {Planelles}, {Potter}, {Quilis}, {Rasera}, {Ricker}, {Roy}, {Ruiz},
  {Sgro}, {Springel}, {Stadel}, {Sutter}, {Tweed}, \&
  {Zemp}}]{knebe_halo_overview_2013}
{Knebe} A., {Pearce} F.~R., {Lux} H., {Ascasibar} Y., {Behroozi} P., {Casado}
  J., {Corbett Moran} C., {Diemand} J., {Dolag} K., {Dominguez-Tenreiro} R.,
  {Elahi} P., {Falck} B., {Gottloeber} S., {Han} J., {Klypin} A., {Lukic} Z.,
  {Maciejewski} M., {McBride} C.~K., {Merchan} M.~E., {Muldrew} S.~I.,
  {Neyrinck} M., {Onions} J., {Planelles} S., {Potter} D., {Quilis} V.,
  {Rasera} Y., {Ricker} P.~M., {Roy} F., {Ruiz} A.~N., {Sgro} M.~A., {Springel}
  V., {Stadel} J., {Sutter} P.~M., {Tweed} D., {Zemp} M., 2013{\natexlab{b}},
  submitted to MNRAS, ArXiv:1304.0585

\bibitem[{Knollmann \& Knebe(2009)}]{knollmann_ahf:_2009}
Knollmann S.~R., Knebe A., 2009, ApJS, 182, 608

\bibitem[{{Komatsu} {et~al.}(2011){Komatsu}, {Smith}, {Dunkley}, {Bennett},
  {Gold}, {Hinshaw}, {Jarosik}, {Larson}, {Nolta}, {Page}, {Spergel},
  {Halpern}, {Hill}, {Kogut}, {Limon}, {Meyer}, {Odegard}, {Tucker}, {Weiland},
  {Wollack}, \& {Wright}}]{komatsu_etal_2011}
{Komatsu} E., {Smith} K.~M., {Dunkley} J., {Bennett} C.~L., {Gold} B.,
  {Hinshaw} G., {Jarosik} N., {Larson} D., {Nolta} M.~R., {Page} L., {Spergel}
  D.~N., {Halpern} M., {Hill} R.~S., {Kogut} A., {Limon} M., {Meyer} S.~S.,
  {Odegard} N., {Tucker} G.~S., {Weiland} J.~L., {Wollack} E., {Wright} E.~L.,
  2011, ApJS, 192, 18

\bibitem[{{Kuhlen} {et~al.}(2012){Kuhlen}, {Vogelsberger}, \&
  {Angulo}}]{Kuhlen_review_2012}
{Kuhlen} M., {Vogelsberger} M., {Angulo} R., 2012, Physics of the Dark
  Universe, 1, 50

\bibitem[{{Lacey} \& {Cole}(1993)}]{lacey_and_cole_1993}
{Lacey} C., {Cole} S., 1993, MNRAS, 262, 627

\bibitem[{{Lee} \& {Yi}(2013)}]{lee_2013}
{Lee} J., {Yi} S.~K., 2013, ApJ, 766, 38

\bibitem[{{Libeskind} {et~al.}(2011){Libeskind}, {Knebe}, {Hoffman},
  {Gottl{\"o}ber}, \& {Yepes}}]{Libeskind_2011}
{Libeskind} N.~I., {Knebe} A., {Hoffman} Y., {Gottl{\"o}ber} S., {Yepes} G.,
  2011, MNRAS, 418, 336

\bibitem[{{Onions} {et~al.}(2012){Onions}, {Pearce}, {Lux}, {Muldrew}, {Knebe},
  \& S.}]{onions_2012}
{Onions} J., {Pearce} F., {Lux} H., {Muldrew} S., {Knebe} A., S. K., 2012,
  submitted to MNRAS, arXiv:1212.0701

\bibitem[{{Roukema} {et~al.}(1997){Roukema}, {Quinn}, {Peterson}, \&
  {Rocca-Volmerange}}]{roukema_etal_1997}
{Roukema} B.~F., {Quinn} P.~J., {Peterson} B.~A., {Rocca-Volmerange} B., 1997,
  MNRAS, 292, 835

\bibitem[{{Skibba} \& {Sheth}(2009)}]{Skibba_2009}
{Skibba} R.~A., {Sheth} R.~K., 2009, MNRAS, 392, 1080

\bibitem[{Springel(2005)}]{springel_cosmological_2005}
Springel V., 2005, MNRAS, 364, 1105

\bibitem[{{Springel} {et~al.}(2005){Springel}, {White}, {Jenkins}, {Frenk},
  {Yoshida}, {Gao}, {Navarro}, {Thacker}, {Croton}, {Helly}, {Peacock}, {Cole},
  {Thomas}, {Couchman}, {Evrard}, {Colberg}, \&
  {Pearce}}]{springel_millennium_2005}
{Springel} V., {White} S.~D.~M., {Jenkins} A., {Frenk} C.~S., {Yoshida} N.,
  {Gao} L., {Navarro} J., {Thacker} R., {Croton} D., {Helly} J., {Peacock}
  J.~A., {Cole} S., {Thomas} P., {Couchman} H., {Evrard} A., {Colberg} J.,
  {Pearce} F., 2005, Nat, 435, 629

\bibitem[{{Tweed} {et~al.}(2009){Tweed}, {Devriendt}, {Blaizot}, {Colombi}, \&
  {Slyz}}]{Tweed2009}
{Tweed} D., {Devriendt} J., {Blaizot} J., {Colombi} S., {Slyz} A., 2009, A\&A,
  506, 647

\end{thebibliography}
\bibliographystyle{mn2e} \label{sec:Bibliography}

\label{lastpage}
\appendix

\section{The tree data format}\label{sec:treeformat}

\begin{table*}
\caption{The {\sc ascii} data format that participants were asked to
  use to return their merger tree results.}
\label{tab:treeformat}
\begin{tabular}{ll}
\hline
Information to be returned& Notes\\
\hline
FormatVersion & $=1$ -- an integer indicating the format version\\
Description & Name of code, version/date of generation; max 1024 characters\\
Nhalo & Total number of \halos\ specified in this file\\
HaloID$_1$, $N_1$ & Halo's ID  and number of direct progenitors\\
Progenitor$_1$ & Halo ID of main progenitor of halo HaloID$_1$ (where $N_1>0$)\\
Progenitor$_2$ & Halo IDs of other progenitors of halo HaloID$_1$\\
\ldots & \ldots\\
Progenitor$_{N_1}$ & Halo ID of last progenitor of halo HaloID$_1$\\
\\
\ldots& \ldots\\
\\
HaloID$_\mathrm{Nhalos}$, $N_\mathrm{NHalo}$ & Halo's ID  and number of direct progenitors\\
Progenitor$_\mathrm{NHalo}$ & Halo ID of main progenitor of halo HaloID$_\mathrm{NHalo}$ (where $N_\mathrm{NHalo}>0$)\\
Progenitor$_2$ & Halo IDs of other progenitors of halo HaloID$_\mathrm{NHalo}$\\
\ldots & \ldots\\
Progenitor$_{N_\mathrm{NHalo}}$ & Halo ID of last progenitor of halo HaloID$_\mathrm{NHalo}$\\
END & String 'END' indicating the last line of the output file\\
\hline
\end{tabular}
\end{table*}

In order to facilitate comparison and use of merger tree data, it is
our intention to define in a future paper a common merger tree
data format.  This should make provision for: required minimal data to
define a merger tree; desired fields to ease use; and the ability to
include optional additional data that may prove useful.  At the time
of writing (prior to the {\sc Sussing Merger Trees} Workshop)
that format had not been defined and so we restrict ourselves to
outlining here the minimal data format that was used for the work
described in this paper.

We supplied each participant in the tree comparison project with a
list of \halos, together with their properties (as described in
Section~\ref{sec:halos}) and an inclusive list of particle IDs.  Each
halo had a identifier (halo ID) that was unique across snapshots.

We required participants to return their results in the {\sc ascii} format
described in Table~\ref{tab:treeformat}, where there is an entry for
each halo.  That contains enough information for us to be able to
reconstruct the merger trees and, in conjuntion with the original halo
list, to follow the growth of \halos over time.

\end{document}